\documentclass[11pt,twoside]{article}
\usepackage{atmp,latexsym}
\copyrightnotice{2003}{7}{205}{232}       %year, volume, first page, last page.

\def\ipo{\hbox{\bf 0}} 
\let\top=\Diamond
\def\zm{z_{\rm min}}
\def\sst#1{{_{#1}}}
\def\xx{}  \def\yy{/}
\catcode`\"=\active \let"=\"

\def\ifundefined#1{\expandafter\ifx\csname#1\endcsname\relax}
\def\bye{\end{document}}   
\long\def\new#1\endnew{{\bf #1}}
\long\def\del#1\enddel{} 

\def\HS#1 {\hspace*{#1pt}} \def\VS#1 {\vspace*{#1pt}}
\def\BC{\begin{center}}    
\def\EC{\end{center}}

\let\0=\over      \let\1=\vec      \def\2{{1\over2}}    \let\3=\ss
\let\4=\underline \let\5=\overline \let\6=\partial      \def\7#1{{#1}\llap{/}}
\def\8#1{{\textstyle{#1}}}         \def\9#1{{\ifmmode{\pmb{#1}}\else\bf#1\fi}}

          \def\({\left(}       \def\){\right)}

\def\eeql#1 {\label{#1}\end{equation}}      \let\nn=\nonumber  
\def\beq{\begin{equation}}      \def\eeq{\end{equation}}        
\def\bea{\begin{eqnarray}}      \def\eea{\end{eqnarray}}

\let\and=\wedge

\let\bra=\langle        \let\ket=\rangle        \def\<#1\>{\bra #1 \ket}
\let\ni=\noindent

\def\rel#1 #2{\buildrel #1 \over {#2}}  

   \let\l=\lambda  \let\m=\mu      
\let\n=\nu            \let\p=\pi

             \let\D=\Delta

\def\IR{{\mathbb R}} \def\IC{{\mathbb C}} \def\IP{{\mathbb P}} 
   
\def\IZ{{\mathbb Z}}

\def\plb#1 #2 {Phys. Lett. {\bf B#1} #2 }
\def\phr#1 #2 {Phys. Rep. {\bf  #1} #2 }        
\def\npb#1 #2 {Nucl. Phys. {\bf B#1} #2 }
\def\atmp#1 {Adv. Theor. Math. Phys. {\bf #1} }
\begin{document}

\title{Affine Kac-Moody algebras, CHL strings and the classification of tops}

\url{hep-th/0303218}         %Everything after http://xxx.lanl.gov/

\author{Vincent Bouchard and Harald Skarke}          %Can have multiple \author, \address
%\author{Harald Skarke}
\address{Mathematical Institute, University of Oxford,\\
24-29 St. Giles', Oxford OX1 3LB, England}
\addressemail{bouchard@maths.ox.ac.uk, skarke@maths.ox.ac.uk}     %Under \address only, please!
%\addressemail{skarke@maths.ox.ac.uk}     %Under \address only, please!
\markboth{\it 
Affine Kac-Moody algebras, CHL strings and the classification of tops}
{\it Vincent Bouchard, Harald Skarke}

\begin{abstract}
Candelas and Font introduced the notion of a `top' as half of a three 
dimensional reflexive polytope and noticed that Dynkin diagrams of enhanced
gauge groups in string theory can be read off from them.
We classify all tops satisfying a generalized definition as a lattice polytope
with one facet containing the origin and the other facets at distance one
from the origin.
These objects torically encode the local geometry of a 
degeneration of an elliptic fibration.
We give a prescription for assigning an affine, possibly twisted Kac-Moody 
algebra to any such top (and more generally to any elliptic fibration 
structure) in a precise way that involves the lengths of simple roots 
and the coefficients of null roots.
Tops related to twisted Kac-Moody algebras can be used to construct string
compactifications with reduced rank of the gauge group.
\end{abstract}

\vfill
\newpage

\section{Introduction}
%\enddel
Toric geometry and Batyrev's construction \cite{Ba} provide a very useful 
setup to study dualities between heterotic strings compactified on a 
Calabi-Yau $n$-fold and F-theory (or type II) compactified on a Calabi-Yau 
$(n+1)$-fold. 
In \cite{CF}, Candelas and Font used reflexive polyhedra to study the 
conjectured duality between the $E_8 \times E_8$ heterotic string 
compactified on the manifold 
$K3 \times T^2$ and the IIA string compactified on a Calabi-Yau threefold. 
It was noticed, and later explained in \cite{egs}, that the affine Dynkin 
diagrams of nonabelian gauge groups occurring in type IIA and also in F-theory 
can be read off from the dual reflexive polyhedron corresponding to the 
Calabi-Yau manifold used for compactification. 
The fibration stucture of the Calabi-Yau 
manifold can be directly seen as a nesting structure of the 
reflexive polyhedron. 
The elliptic fibration structure of the $K3$ part of the 
Calabi-Yau divides the three-dimensional reflexive polyhedron corresponding 
to the $K3$ in two parts, a top and a bottom, separated by the 
two-dimensional reflexive polygon of the fiber. 
The concept of top was then introduced as half of a reflexive polyhedron.

This was just the beginning of the story. 
The ideas of \cite{CF} were studied in detail in many other papers 
\cite{CPR1,CPR2,CPR3,CPR4,fst,BM,HLY}. 
The Calabi-Yau manifolds analysed in these papers were $K3$ fibrations 
with an elliptically fibered $K3$ manifold where the elliptic fibration 
structure of the $K3$ carries over to the Calabi-Yau manifold. 
These nested fibration structures can be seen explicitly in the toric 
diagrams as nestings of the corresponding reflexive polyhedra.
The fan for the toric variety describing the base of the elliptic fibration 
is given by projecting the higher dimensional fan corresponding to the fibered 
Calabi-Yau manifold along the two dimensions of the reflexive polygon that 
represents the fiber \cite{fft}.
Then the way the elliptic fibers degenerate along the curves in the base 
space can be found, in order to determine the enhanced gauge groups, by 
considering the preimage of the projection for each toric divisor in the base. 
The concept of top can now be generalized \cite{fst} to the geometrical 
objects formed by the preimages of the corresponding toric divisors. 
These objects are three-dimensional lattice polyhedra 
with one facet containing the origin and the other facets
at integral distance one from the origin. 
This definition implies that the facet containing the origin is a reflexive 
polygon. 
Note that this really generalizes the concept initially introduced by Candelas 
and Font, since the tops defined as half of a reflexive polyhedron have all 
the properties of the new tops, but the new tops cannot always be completed 
to reflexive polyhedra.
Alternatively the more general definition can be seen as the description 
of a toric hypersurface that is an elliptic fibration over $\IC$ and 
its degeneration over $0$.

In this paper we classify all the possible tops, using the general definition.
In contrast to the case of reflexive polyhedra, there are 
infinitely many tops, even for each choice of one of the 16 reflexive polygons
as the facet containing the origin.
We find that there is a precise prescription for assigning an affine Kac-Moody 
algebra to any top, in a way that involves the lengths of simple roots and the 
coefficients of the null root.
Owing to this fact the classification of tops is related to that of affine 
Kac-Moody algebras. 
We also find one parameter families as well as sporadic cases.
In addition,
for each of the 16 polygons there is also a family depending on $l-3$ 
integer parameters where $l$ is the number of lattice points of the polygon;
these correspond to the $A_n^{(1)}$ series of affine Kac-Moody algebras in 
such a way that $n$ is a linear combination of the parameters.
Each of the untwisted affine Kac-Moody algebras occurs quite 
a number of times, and in addition four of the six possible (families of)
twisted algebras also occur.
The tops featuring the latter are related in a very nice way to string 
compactifications with reduced rank, i.e. CHL strings \cite{CHL} and their 
generalisations and duals.

We have structured the paper in such a way that it reflects the hierarchy
\beq \hbox{Lattice polytopes}\quad\to\quad
{\hbox{Toric geometry}\quad\atop\hbox{Kac-Moody algebras}}
\quad\to\quad\hbox{String theory.}\nonumber\end{equation}
So it is possible to read only sections 2 and 4 as an exercise in lattice
polytopes, and to read everything except the second half of section 5 
without knowledge of string theory.
Section 2 contains basic facts about reflexive polytopes, tops and their 
duals.
Section 3 gives interpretations of these concepts in terms of toric geometry
and a general argument, based on Kodaira's classification of degenerations of
elliptic fibrations, that not only tops but also elliptic fibration structures 
in general should be related to untwisted or twisted affine Kac-Moody algebras.
In section 4 we present our classification scheme.
The last section gives a discussion of our results with particular emphasis
on the cases related to twisted algebras, first in terms of geometry (twisted 
algebras occur only for fibrations that allow orbifold actions) and then in 
terms of dualities between M-theory, F-theory or type II strings and heterotic 
strings or CHL type strings.
Finally there is an appendix containing the results of the classification.

\vfill
\newpage
\section{Reflexive polytopes, tops and their duals}

Throughout this paper we will work a dual pair of lattices $M\simeq \IZ^3$
and $N\simeq \IZ^3$, view them as subsets of vector spaces 
$M_\IR=M\otimes_\IZ\IR\simeq \IR^3$ and 
$N_\IR=N\otimes_\IZ\IR\simeq \IR^3$, and denote the pairings 
$M\times N\to \IZ$ and $M_\IR\times N_\IR \to
\IR$ by $(u,v)\to \<u,v\>$.

A polytope in $M_\IR$ is the convex hull of a finite number of points in
$M_\IR$, and a polyhedron in $M_\IR$ is the intersection of finitely
many half-spaces (given by inequalities $\<u,v\>\ge c$ with some $v\in
N_\IR$ and $c\in \IR$) in $M_\IR$.
It is well known that any polytope is a polyhedron and any bounded
polyhedron is a polytope.
If a polyhedron $S\subset M_\IR$ contains the origin \ipo, its dual 
\beq S^*=\{v\in N_\IR : \<u,v\>\geq -1 \hbox{ for all } u \in
S\}. \end{equation}
is also a polyhedron containing \ipo, and $(S^*)^*=S$.

A lattice polytope in $M_\IR$ is a polytope with vertices in $M$.
A polytope $\D\subset M_\IR$ containing \ipo\ is called reflexive if
both $\D$ and $\D^*$ are lattice polytopes.
This is equivalent to $\D$ being a lattice polytope whose bounding
equations are of the form $\<u,v_i\>\ge -1$ with $v_i\in N$
(in coordinates, $\sum_j u_j v_{ij}\ge -1$ with integer coefficients
$v_{ij}$). 
By convexity it is sufficient to consider only those equations
corresponding to $v_i$ that are vertices of $\D^*$.
In this way there is a duality between vertices of $\D^*$ and facets
of $\D$; similarly, there are dualities between $p$-dimensional faces
of $\D$ and $(n-p-1)$-dimensional faces of $\D^*$ (in three dimensions:
between edges and dual edges).

An interior point $u$ of a reflexive polytope must satisfy 
$\<u,v_i\> > -1$ for all $v_i$, so an interior lattice point must
satisfy $\<u,v_i\>\ge 0$. 
Thus if $u$ is an interior lattice point, then $nu$ is also an
interior lattice point for any nonnegative integer $n$.
For $u\ne$ \ipo\ this would be in conflict with the boundedness of
$\D$, implying that \ipo\ is the only interior lattice point.

In \cite{CF} Candelas and Font considered reflexive polytopes whose
intersections with a plane were themselves reflexive polygons;
the fact that this intersection cuts the polytope into two parts
(`top' and `bottom') gave rise to the notion of a `top' as
half of a reflexive polytope in this sense.
In \cite{fst} this definition was generalized in the following way.
A top $\top\subset N_\IR$ is a lattice polytope such that one of its
defining inequalities is of the form $\<u_0,v\>\ge 0$ and all others
are of the form $\<u_i,v\>\ge -1$, with $u_i\in M$.

We consider two tops to be isomorphic if they are related by a
$GL(3,\IZ)$ transformation.
This allows us to choose coordinates $(x,y,z)$ for
$M$ and $M_\IR$ such that $u_0$ has coordinates $(0,0,1)$
(we will always make this choice whenever we work with specific coordinates).
Then the inequality corresponding to the facet 
$F_0:=\{v\in \top:\<u_0,v\>=0\}$ is given by $\bar z\ge 0$ in terms of
dual coordinates $(\bar x,\bar y,\bar z)$ for $N_\IR$.
$F_0$ is bounded by the restrictions of the other inequalities to
$\bar z=0$; as these are again of the type $\ldots \ge -1$ with
integer coefficients, $F_0$ is a reflexive polygon.
Thus the more general definition of a top indeed contains all the
cases of \cite {CF}.
A straightforward adaptation of the above argument about reflexive
polytopes shows that a top has no interior lattice points.

The dual $\top^*\subset M_\IR$ of $\top$ is the polyhedron defined by
the inequalities originating from the vertices of $\top$.
The vertices $(\bar x_i,\bar y_i,0)$ of
$F_0$ lead to inequalities of the form $x\bar x_i+y\bar y_i\ge -1$;
we will refer to the corresponding facets as `vertical facets'.
Thus $\top^*$ must be contained in a prism over $F_0^*$ (the dual of
$F_0$ in the two dimensional sense).
The remaining vertices of $\top$ have $\bar z>0$.
The corresponding inequalities can be written as 
\beq z\bar z_i\ge -1-x\bar x_i-y\bar y_i, \end{equation}\label{nonvert} 
implying that for
every fixed $(x,y)\in F_0^*$ there is a minimal (but no maximal) value
$z_{\rm min}(x,y)$ such that $(x,y,z)\in \top^*$ for all 
$z \ge z_{\rm min}(x,y)$.
In this way we can view $\top^*$ as the result of `chopping off' the
lower parts of an infinitely extended prism.
Alternatively, we may see it as a `polytope' with one vertex
$u_\infty=+\infty\, u_0$ at infinity, as it is the dual of $\top$ which
may be defined by $\<u_i,v\>\ge -1$ for $i\ge 1$ and 
$\<\l u_0,v\>\ge -1$ for arbitrarily large positive $\l$.
$\top^*$ has infinitely many interior lattice points $(0,0,z)$ with
$z$ any nonnegative integer.

The projection 
\beq \pi : \;\;\top^* \to F_0^*,\;\;(x,y,z)\to(x,y)  \end{equation}
takes vertices of $\top^*$ to lattice points of $F_0^*$.
Conversely, every (finite) vertex of $\top^*$ is of the form
$(x,y,\zm(x,y))$ where $(x,y)$ is a lattice point of $F_0^*$.
This means that by specifying $F_0^*$ and $z_{\rm min}$ for each of
its lattice points, we have specified $\top^*$ completely.
The projections of non-vertical facets determine a partition of
$F_0^*$.
These facts afford a useful pictorial representation of $\top^*$ in
terms of a picture of $F_0^*$ where every lattice point is labeled
with the corresponding $z_{\rm min}$.

\begin{figure}[!hbt]
\begin{center}
\includegraphics[width=11cm]{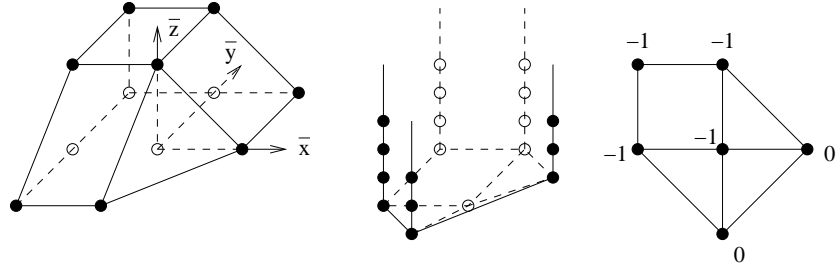}
\caption{A top, its dual and the minimal point notation.}\label{fig:a3}
\end{center}
\end{figure}

Let us now work out an explicit example. 
Suppose $\top$ is the convex hull of $(-1,-1,0)$, $(-1,1,0)$,
$(1,1,0)$, $(1,0,0)$, $(0,-1,0)$, $(0,0,1)$, $(-1,0,1)$, $(-1,1,1)$
and $(0,1,1)$, as shown in the first part of figure~\ref{fig:a3}. 
$\top$ has 7 facets corresponding to the inequalities
\bea
&&F_0:\;\bar z\ge 0, \quad F_1:\;\bar z\le 1, \quad F_2:\;\bar x\ge -1, 
\quad F_3:\;\bar y\le 1,\nn\\
&&F_4:\;\bar x+\bar z\le 1, \quad F_5:\;\bar x-\bar y+\bar z\le 1, 
\quad F_6:\;\bar z-\bar y\le 1, 
\nn\eea
implying that $\top$ is indeed a top and that $\top^*$ has the
vertices $(0,0,-1)$, $(1,0,0)$, $(0,-1,0)$, $(-1,0,-1)$, $(-1,1,-1)$,
$(0,1,-1)$, in addition to the `vertex at infinity' $(0,0,\infty)$.
Thus $F_0^*$ is the convex hull of $(1,0)$, $(0,-1)$, $(-1,0)$, $(-1,1)$,
$(0,1)$ with $z_{\rm min}$ as shown in the third diagram of
figure~\ref{fig:a3}. 

%\del

\section{Toric interpretation and affine Kac-Moody algebras}

In this section we are going to intepret the concepts introduced in the 
previous one in terms of toric geometry \cite{Fu,Oda}, using the 
homogeneous coordinate approach of \cite{Cox}.
An elementary introduction in a similar setup to the background required here
can be found in \cite{sdt}.
We will also find an intriguing connection with affine Kac-Moody algebras.

Given a three dimensional pair of reflexive polytopes $\D\in M_\IR$,
$\D^*\in N_\IR$, a smooth K3 surface can be constructed in the
following manner. 
Any complete triangulation of the surface of $\D^*$ defines a fan
whose three dimensional cones are just the cones over the regular
(i.e., lattice volume one) triangles. 
To any lattice point $p_i=(\bar x_i, \bar y_i, \bar z_i)$ on the
boundary of $\D^*$ one can assign a 
homogeneous coordinate $w_i\in \IC$, with the rule that several $w_i$
are allowed to vanish simultaneously only if there is a cone such that
the corresponding $p_i$ all belong to this cone.
The equivalence relations among the homogenous coordinates are given
by
\beq (w_1,\ldots, w_n) \sim (\l^{k_1}w_1,\ldots,\l^{k_n}w_n)
   \;\hbox{ for any }\l\in\IC\setminus\{ 0\} \end{equation}\label{equiv}
with any set of integer $k_i$ such that $\sum k_i p_i=0$; 
among these relations, $n-3$ are independent.
This construction gives rise to a smooth compact three dimensional toric
variety $V$ (smooth because the generators of every cone are also
generators of $N$, compact because the fan fills $N_\IR$).
The loci $w_i=0$ are the toric divisors $D_i$.

To any lattice point $q_j$ of $M$ we can assign a monomial
$m_j=\prod_i w_i^{\<q_j,p_i\>+1}$; the exponents are nonnegative as a
consequence of reflexivity.
The hypersurface defined by a generic polynomial $P=\sum a_j m_j$,
which transforms homogeneously under (\ref{equiv}), can be shown to define
a K3 hypersurface in $V$.

Suppose the intersection of $\D^*$ with the plane $\bar z = 0$ gives a
reflexive polygon.
We may reinterpret $P$ as a polynomial in the $w_i$ for which 
$\bar z_i=0$, with coefficients depending on the remaining $w_i$, 
i.e. we are dealing with an elliptic curve parametrized by the $w_i$ 
for which $\bar z_i\ne 0$.
The map $V \to \IP^1$,
\beq  (w_1,\ldots, w_n) \;\to \; W=\prod_{i:\bar z_i\ne 0} w_i^{\bar z_i} 
\end{equation}\label{projmap}
is easily checked to be consistent with (\ref{equiv}) and thus well
defined.
At any point of the $\IP^1$ that is neither $0$ nor $\infty$
all the $w_i$ with $\bar z_i\ne 0$ are non-vanishing, and
(\ref{equiv}) can be used to set all except one of them to 1.
This gives the K3 surface the structure of an elliptic fibration.

The situation of a top is very similar.
Here $P=0$ defines a non-compact surface $\cal M$ with the structure of an
elliptic fibration, the main change being that the base space is
now $\IC$ as there are no negative exponents in (\ref{projmap}).
% parametrized by $W=\prod_{i:\bar z_i>0} w_i^{\bar z_i}$.
In those cases where the top is half of a reflexive polytope, it
encodes the geometry of the K3 away from the preimage of the point
$\infty$. 
In addition we have an interpretation of a top in the case of an 
elliptically fibered higher dimensional Calabi--Yau hypersurface in a 
toric variety.
Here the polygon encoding the elliptic fiber is again an intersection 
of a reflexive polytope with a plane.
The base space of the fibration is determined by projecting the fan along
the two dimensions spanned by the polygon \cite{fft}.
Rays in this projected fan determine divisors in the base along which the 
fiber can degenerate; the inverse image of such a ray is again a top whose
structure determines the generic type of degeneration over the intersection 
of a disc with the divisor.
This may lead to tops with far more points than in a three 
dimensional reflexive polytope \cite{fst}.

All of the interesting geometry happens at $W=0$.
Let us start with a top that has only a single vertex $p_n$ at $\bar z
=1$ and all other lattice points at $\bar z =0$.
This leads to a hypersurface determined by a polynomial in
$w_1,\ldots, w_{n-1}$ with coefficients that are power series in $W=w_n$
that start with a constant; each of these power series corresponds to
a vertical edge in $\top^*$.
In the generic case nothing special happens and we get a smooth elliptic
curve at $W=0$.
If we restrict some of the $W$ dependent coefficients to start at
higher powers of $W$, this may lead to singularities.
Now the non-vanishing coefficients correspond only to a subset 
${\top'}^*$ of $\top^*$, and we can resolve the
singularity by passing from $\top$ to $\top'$, which corresponds
to a blow up.

\begin{figure}[h]
\begin{center}
\includegraphics[width=127mm]{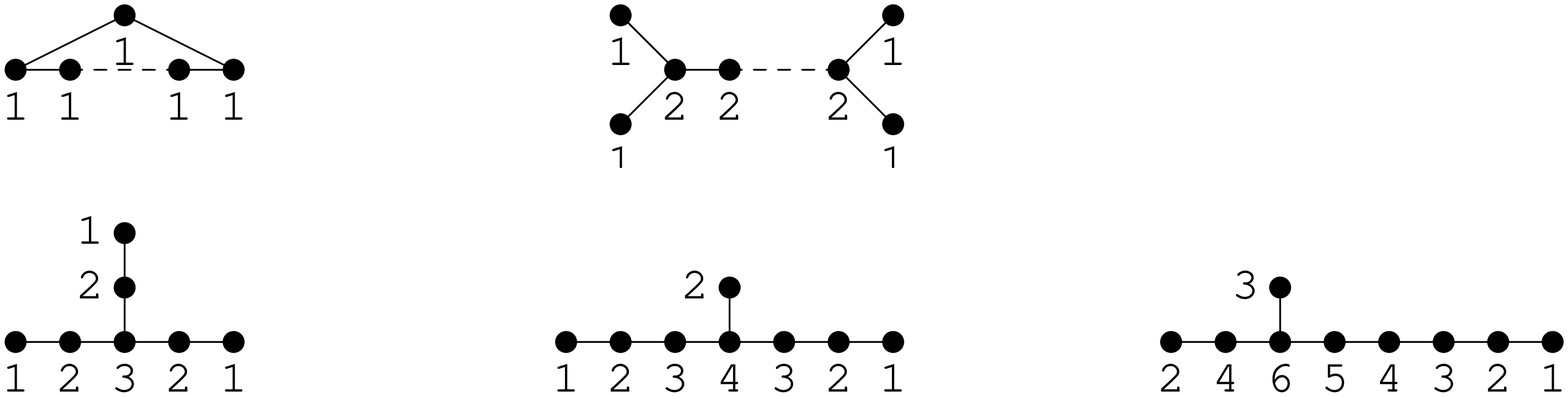}
\put(-350,100){$A_n^{(1)}$}
\put(-200,100){$D_n^{(1)}$}
\put(-360,40){$E_6^{(1)}$}
\put(-220,40){$E_7^{(1)}$}
\put(-55,40){$E_8^{(1)}$}
\caption{Dynkin diagrams of the self-dual untwisted ADE Kac-Moody algebras.}
\label{fig:ADE}
\end{center}
\end{figure}

An arbitrary top $\top$ always contains at least one lattice point at
$\bar z=1$. 
This can be seen by observing that a complete triangulation of the fan
leads to a triangulation of $\top$ in terms of tetrahedra of volume
one; such a tetrahedron with base at $\bar z=0$ must then have its
apex at $\bar z =1$.
Thus every top can be interpreted as the smooth resolution of the
singularity at $W=0$ of an elliptic fibration.
Such singularities and their resolutions were classified by Kodaira
\cite{Kod}, resulting in the following picture (see also \cite{BPV}).
Under certain assumptions fulfilled in the present case, the inverse
image of $W=0$ must consist either of a single (possibly singular)
curve or of a collection of smooth rational curves $C_i$ such that the
intersections of these curves obey $C_i\cdot C_j=-M_{ij}$ where $M$ is
the Cartan matrix of an untwisted Kac-Moody algebra of ADE type 
(these are precisely the self-dual affine Kac-Moody algebras); the
multiplicities of the $C_i$ are the coefficients of the null vector of
the algebra.
In other words, the intersection patterns and multiplicities are
respresented by the Dynkin diagrams with labels, as shown in
figure~\ref{fig:ADE}. 

\begin{figure}[h]
\begin{center}
\includegraphics[width=10cm]{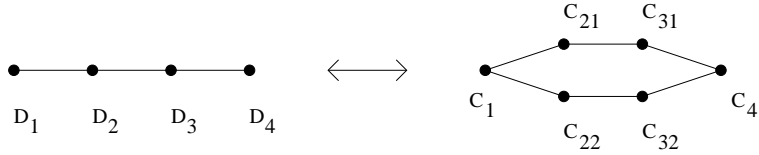}
\end{center}
\caption{An edge of $\top^*$ and the intersection pattern to which it corresponds.}
\label{fig:inters}
\end{figure}

Every toric divisor $D_i=\{w_i=0\}\subset V$ may give rise via 
$D_i\cdot {\cal M}=\sum_j C_{ij}$ to one or more curves $C_{ij}$ in $\cal M$.
This type of intersection theory was worked out in detail for three 
dimensional reflexive polytopes in \cite{egs}, with the following results 
which also hold in the context of tops for the compact divisors with 
$\bar z > 0$.
A lattice point interior to a facet of $\top$ determines a divisor
$D_i$ that does not intersect $\cal M$.
A vertex gives rise to a single curve (rational for any vertex at
$\bar z>0$ in a non-trivial top).
A point interior to an edge corresponds to $l$ 
rational curves, where $l$ is the length of the dual edge.
Each of the rational curves involved has self intersection $-2$.
Mutual intersections occur only among neighbours along edges; in that
case $D_i\cdot D_j \cdot {\cal M}=l$.
For example, an edge of length 3 dual to an edge of length 2 gives
rise to an intersection pattern as indicated in figure~\ref{fig:inters}.
Here $D_1$ and $D_4$ come from vertices $v_1,v_4$ of $\top$ and $D_2$ and
$D_3$ from points interior to the edge $\5{v_1v_4}$.
$D_2$ and $D_3$ each give rise to two curves ($C_{21}, C_{22}$ and
$C_{31}, C_{32}$, respectively) in such a way that any two curves have
mutual intersection one if they are joined by an edge in the second
part of figure~\ref{fig:inters}.

We can use this information to predict the structure of the edge
diagram of the part of $\top$ with $\bar z \ge 1$.
If all the dual edges have length 1, it must have the structure of the
Dynkin diagram of an affine ADE algebra.
From (\ref{projmap}) it is clear that the multiplicity of a curve
$C_i=D_i\cdot {\cal M}$ in $W=0$ is just $\bar z_i$, i.e. the Dynkin
labels encode the heights $\bar z$ of the lattice points.
If some of the dual edges have lengths $>1$, the edge diagram must be
the result of partially collapsing an ADE diagram,
as in reading figure~\ref{fig:inters} from right to left.
Each of the curves $C_{ij}$ originating from the same $D_i$ must have
multiplicity $\bar z_i$ in $W=0$.
There are two possibilities. 
If both vertices are at $\bar z > 0$, the uncollapsed diagram contains
a closed loop and the only possibility is the $A$-series.
\begin{figure}[h]
\begin{center}
\includegraphics[width=127mm]{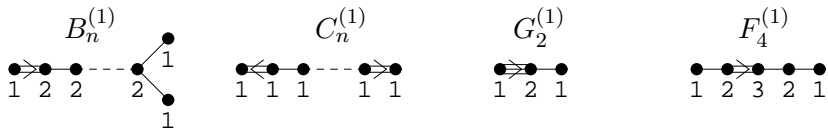}
\put(-320,40){$B_n^{(1)}$}
\put(-225,40){$C_n^{(1)}$}
\put(-150,40){$G_2^{(1)}$}
\put(-65,40){$F_4^{(1)}$}
\caption{Dynkin diagrams of the duals of untwisted non-simply laced Kac-Moody 
algebras.}
\label{fig:utnsl}
\end{center}
\end{figure}
The collapsed diagram must then look like the second one in figure
\ref{fig:utnsl}, where we use multiple lines and arrows to indicate
that we pass from a point giving rise to more than one curve to a vertex
associated with a single curve.
The other possibility for an edge whose dual has length $>1$ is that one 
of its vertices is at $\bar z =0$.
Then we
have to identify two or more ends of one of the $D$ or $E$ Dynkin diagrams.
Direct inspection shows that this is possible only for those ends
whose last point has a label of 1.
These are precisely the `extension points' if the ADE diagram is read
as the extended Dynkin diagram of an ADE Lie algebra.
If the folding procedure leaves at least one of these points
invariant, we may view this as folding an ordinary Dynkin diagram,
taking us from a self-dual simply laced algebra to the dual of a
non-simply laced one.
In this way we get $B_n$ from $D_{n+1}$, $C_n$ from $A_{2n-1}$, $G_2$
from $D_4$ (by a triple folding), and $F_4$ from $E_6$.

\begin{figure}[h]
\begin{center}
\includegraphics[width=127mm]{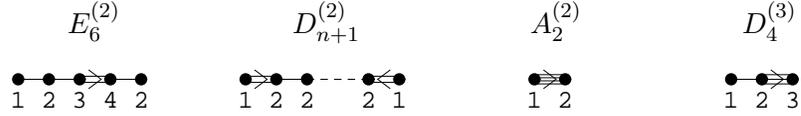}
\put(-320,40){$E_6^{(2)}$}
\put(-235,40){$D_{n+1}^{(2)}$}
\put(-145,40){$A_2^{(2)}$}
\put(-65,40){$D_4^{(3)}$}
\caption{Dynkin diagrams of the duals of twisted Kac-Moody algebras
that can be read off from tops.} 
\label{fig:twist}
\end{center}
\end{figure}
If all of the points with Dynkin label 1 are involved in the folding
procedure, we pass from an untwisted Kac-Moody algebra to a twisted
one.
We find the possibilities $E_7^{(1)}\to E_6^{(2)}$, 
$D_{n+3}^{(1)}\to D_{n+1}^{(2)}$ (two foldings), 
$D_4^{(1)}\to A_2^{(2)}$ (quadruple folding) and 
$E_6^{(1)}\to D_4^{(3)}$ (triple folding); the resulting diagrams are
shown in figure~\ref{fig:twist}.
Our notation is the one used, for example, in \cite{GO}.
Two further twisted algebras $A_{2r-1}^{(2)}$ and
$A_{2r}^{(2)}$ that come from foldings of $D$ diagrams that leave only the
central point invariant cannot occur in the context of tops.

In applications to string theory we are often interested in higher dimensional 
geometries such that locally there is a product structure involving a 
neighborhood of a degeneration of an elliptic fibration.
For example, the total space may be a higher dimensional elliptically fibered 
Calabi--Yau space or a $K3$ bundle over $S^1$.
Then it may happen that there is a closed loop such that over every 
neighborhood in the loop we have one of Kodaira's degenerations, but upon 
going around the loop the exceptional curves get permuted.
Using the fact that the permuted curves intersect if and only if the original 
curves intersect and otherwise the same arguments as before (in particular,
matching of multiplicities), we see that a folding of an affine ADE diagram 
can be assigned in any such case, independently of whether we have a 
description in terms of a top.

Before proceeding to the classification of tops, let us also discuss the toric
interpretation of the vertices of a top $\top$ in the plane $\bar z = 0$.
These are just the vertices of the polygon $F_0$.
In terms of the geometry of the elliptic curve determined by $F_0$ every 
such vertex $v$ gives rise to $l$ divisors (i.e., points) in the elliptic 
curve where $l$ is the length of the edge of $F_0^*$ dual to $v$.
In the context of $\top$ there are $l$ sections of the fibration for generic 
values of the coefficients.
If $\top$ is part of a three or higher dimensional reflexive polytope, 
$v$ determines a divisor $D$ in the corresponding Calabi--Yau hypersurface 
that may be reducible or irreducible.
In the latter case this divisor projects to an $l$-fold cover of the base 
space of the fibration.
In the case of a three dimensional reflexive polytope, $D$ consists of $l$
curves in the corresponding K3 if $v$ is interior to an edge and is 
irreducible if $v$ is a vertex of the three dimensional polytope.

\section{Classification}

\begin{figure}[!tbp]
\begin{center}
\includegraphics[width=12cm]{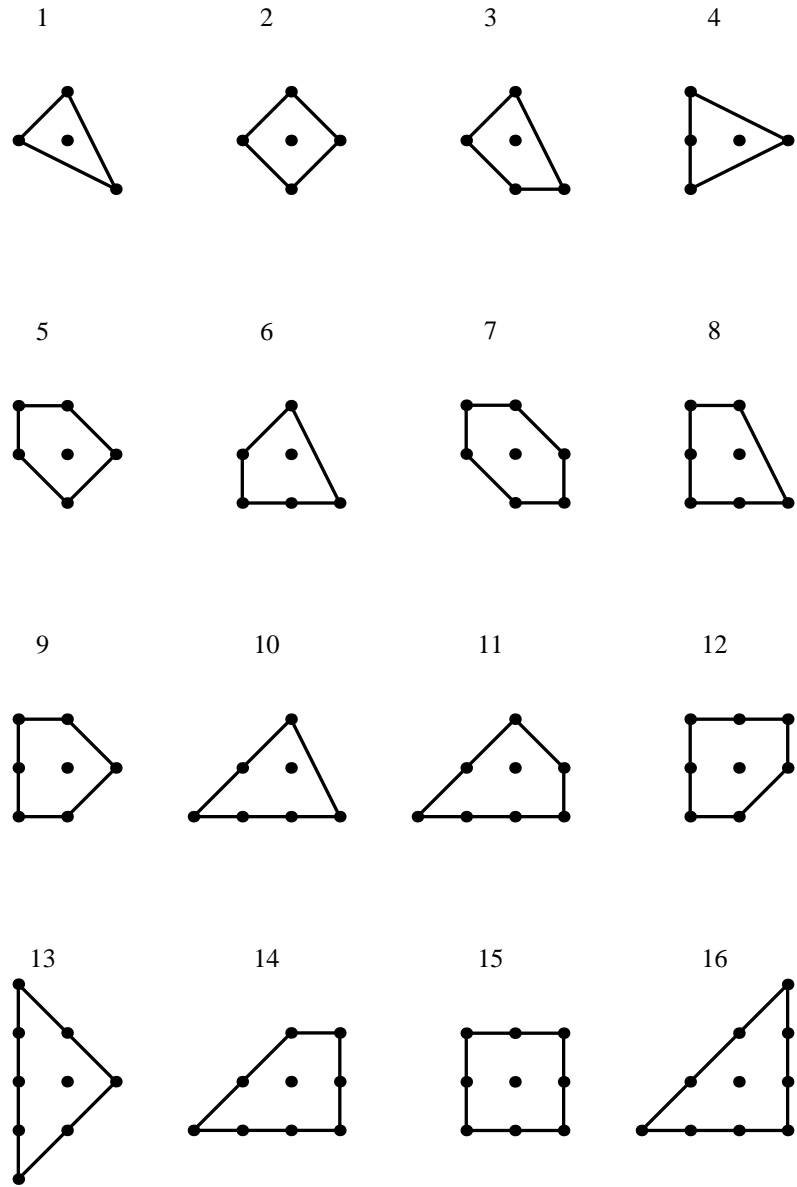}
\end{center}
\caption{The 16 two-dimensional reflexive polygons. 
The polygons $1,2,\ldots,6$ are respectively dual to the polygons
$16,15,\ldots,11$, and the polygons $7,\ldots,10$ are self-dual.}
\label{fig:16pol}
\end{figure}

The classification of reflexive polygons is well known \cite{Ba85,Koe}.
As we make extensive use of it, the resulting 16 polygons are shown in
figure \ref{fig:16pol}.
More recently, a general algorithm for classifying reflexive polytopes
was developed \cite{crp,wtc,rwf} and successfully applied to the 
three \cite{c3d} and four dimensional \cite{c4d} cases.

The main idea of the algorithm of \cite{crp,wtc,rwf} is to look for a set of
maximal polytopes that contain all others; these are dual
to minimal polytopes that do not contain any other reflexive
polytopes.
In the present case of classifying tops, we lose the symmetry between
the objects we are trying to classify and their duals.
Given that the duals are infinite, we will obviously look for maximal
objects among the duals and minimal objects among the tops themselves.
As every dual $\top^*$ of a top must be contained in a prism over one
of the 16 polygons of figure~\ref{fig:16pol}, it is natural to treat
these prisms as the maximal polyhedra.

We have already drastically reduced the $GL(3,\IZ)$ group of
isomorphisms between tops by demanding that $F_0$ lie in the plane
$\bar z =0$.
A further reduction comes from making a specific choice of coordinates
for $F_0$.
The remaining freedom is in the subgroup $G$ of $GL(3,\IZ)$ that fixes 
$F_0$ (not necessarily pointwise).
The elements of $G$ that fix every point of $F_0$ form a normal
subgroup $G_0\simeq \IZ^2$ of $G$; elements of $G_0$ act via
\beq 
(\bar x,\, \bar y,\, \bar z)\to 
(\bar x + a \bar z,\, \bar y +b \bar z,\, \bar z),
\quad (x,\,y,\,z)\to (x,\,y,\,z-ax-by) \quad 
\hbox{with } a,b\in\IZ.
 \end{equation}\label{actg0}
The quotient $G/G_0$ can be identified with the subgroup of
$GL(2,\IZ)$ that fixes $F_0$; as it must take vertices to vertices and
keep the order (up to reversion), $G/G_0$ must be a subgroup of the
dihedral group of order $2n$ of the $n$-gon $F_0$.
The freedom in $G$ can then be eliminated by using (\ref{actg0}) to fix
$z_{\rm min}$ for two lattice points at the boundary of $F_0^*$ and
dealing with the remaining $G/G_0$ freedom by direct inspection.

The boundary point $b_0:=(0,0,z_0)\in \top^*$ below \ipo\ (with
$z_0:=\zm(0,0)$) is invariant under the transformation (\ref{actg0}).
It must belong to one or more facets of the type (\ref{nonvert}), so 
$1/z_0$ must be a negative integer.
Conversely, for every point $p=(\bar x, \bar y, \bar z)\in \top$,
$\bar z$ is invariant under (\ref{actg0}), and $\<p,b_0\>\ge -1$
implies $\bar z \le -1/z_0$. 
The vertices of $\top$ at $\bar z =-1/z_0$ are dual to the facets of
$\top^*$ that contain $b_0$.

\ni
{\bf Lemma:} If $b_0=(0,0,-1)$, then every non-vertical facet of
$\top^*$ contains $b_0$.\\[1mm]
{\it Proof:} $b_0=(0,0,-1)$ implies that $\top$ is bounded by 
$\bar z \le 1$, so any vertex of $\top$ must have either $\bar z =0$,
corresponding to a vertical facet of $\top^*$, or $\bar z=1$,
corresponding to a facet that contains $b_0=(0,0,-1)$.

We are now in a position to enumerate the cases relevant to the
classification. 

\ni
{\bf Case 0:} $\top^*$ has a single non-vertical facet dual to a
vertex of $\top$ at $\bar z = 1$.\\
We can use (\ref{actg0}) to have this vertex at $(0,0,1)$ and thus
$\zm=-1$ everywhere.
This trivial case exists for every choice of $F_0$.

\ni
{\bf Case 1:} $b_0$ is a vertex of $\top^*$.\\
This implies $b_0=(0,0,-1)$, dual to a facet $F_1$ of $\top$
corresponding to $\bar z \le 1$.
According to the lemma, the structure of $\top^*$ is determined by a
partition of $F_0^*$ in the style of cutting a cake.
Equations of non-vertical facets take the form (\ref{nonvert}) with
$\bar z=1$, implying that $\zm(x,y)$ is integer whenever $x$ and $y$
are integer.
We have seen an example in figure~\ref{fig:a3}; this example should
also serve as useful background for the following discussion.

Consider three consecutive lattice points $p_{i-1}, p_i, p_{i+1}$
along the boundary of $F_0^*$.
It is easily checked that they fulfill $p_{i-1}+p_{i+1}=(2-l_i)\,p_i$ where
$l_i$ is the length (in lattice units) of the edge of $F_0$ dual to $p_i$
(with $l_i=0$ if $p_i$ is not a vertex).
The facets of $\top^*$ whose projections contain the triangles
$b_0\,p_{i-1}\,p_i$ and $b_0\,p_i\,p_{i+1}$, respectively, are dual to
vertices $v_{i-1}, v_i\in F_1$, with $v_{i-1}=v_i$ if 
$p_{i-1}, p_i, p_{i+1}$ belong to the projection of a single
non-vertical facet. 
One can calculate that $v_{i-1},v_i$ have lattice distance 
\beq \zm(p_{i-1})+(l_i-2)\zm(p_i)+\zm(p_{i+1})+l_i; \end{equation}\label{latdist}
nonnegativity of this expression is just the local convexity condition.
%By summing these expressions one obtains that t
The circumference in lattice units of the polygon $F_1$ is the sum 
$\sum_i l_i (\zm(p_i)+1)$ of these expressions.

All possible cases can be enumerated by choosing an integer $\zm$ for
every lattice point at the boundary of $F_0^*$, subject to consistency
with convexity, i.e. nonnegativity of (\ref{latdist}) at each lattice
point of $F_0^*$;
to ensure that $b_0$ is a vertex, we also need that at least three of
these expressions are positive.
The freedom in (\ref{actg0}) can be eliminated, for example, by
putting two adjacent boundary points at $z=-1$.

\begin{figure}[h]
\begin{center}
\includegraphics[width=12cm]{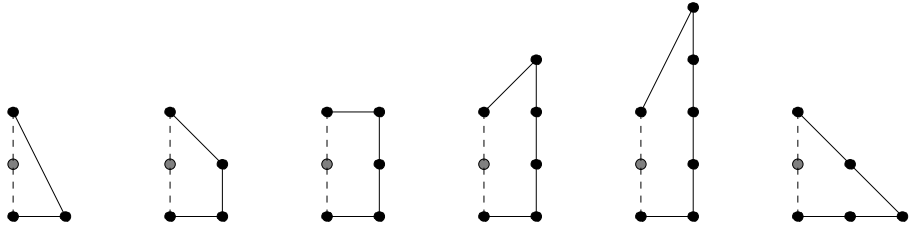}
\caption{Possible halves of reflexive polygons}
\label{fig:dcen}
\end{center}
\end{figure}

\ni
{\bf Case 2:} $b_0$ lies on a line connecting two lattice points $p_1,p_2$
of $\top^*$.\\
Then $b_0=(p_1+p_2)/2$, so $2z_0$ must be integer, i.e. 
$z_0\in\{-1,-1/2\}$.
The projection of $\5{p_1p_2}$ divides $F_0^*$ into two halves.
By inspection of figure~\ref{fig:16pol} we see that either half must
look, up to automorphisms of the two dimensional lattice, like one of
the possibilities shown in figure~\ref{fig:dcen} 
(without loss of generality, we assume that $\5{p_1p_2}$ is at $x=0$).

\ni
{\bf a)} $z_0=-1$: Because of the lemma there are no more than two
non-vertical facets corresponding to the two halves with $x\le 0$ and
$x\ge0$. 
We can use (\ref{actg0}) to put either of them, but not both, at $z=-1$.
If we choose, say, $z=-1$ for the facet at $x\ge0$, then the facet at 
$x\le 0$ must be the transform of a facet at $z=-1$ by (\ref{actg0})
with $b=0$; convexity implies $a\ge 0$.
This gives a one parameter family that starts with the case of a
single facet at $z=-1$ (case 0).
For $a\ge 1$, $\top$ has precisely two vertices at $\bar z=1$ whose
distance is $a$. 

\ni
{\bf b)} $z_0=-1/2$: There are one or two vertices of $\top$ at
$\bar z=2$, dual to the facet(s) containing $b_0$.
All other vertices of $\top$ must be at $\bar z=0$ or $1$.
We can again treat the halves separately and find that after using up
the freedom given by (\ref{actg0}) there are only finitely many cases
for each diagram of figure~\ref{fig:dcen}, corresponding to partitions
such that the facet containing $b_0$ is dual to a vertex at $\bar z=2$
and all other facets correspond to $\bar z=1$.
Here the freedom of applying (\ref{actg0}) to one of the halves leads
to families such that the edge at $\bar z=2$ has length $2a$ or $2a+1$.

\begin{figure}[h]
\begin{center}
\includegraphics[width=12.5cm]{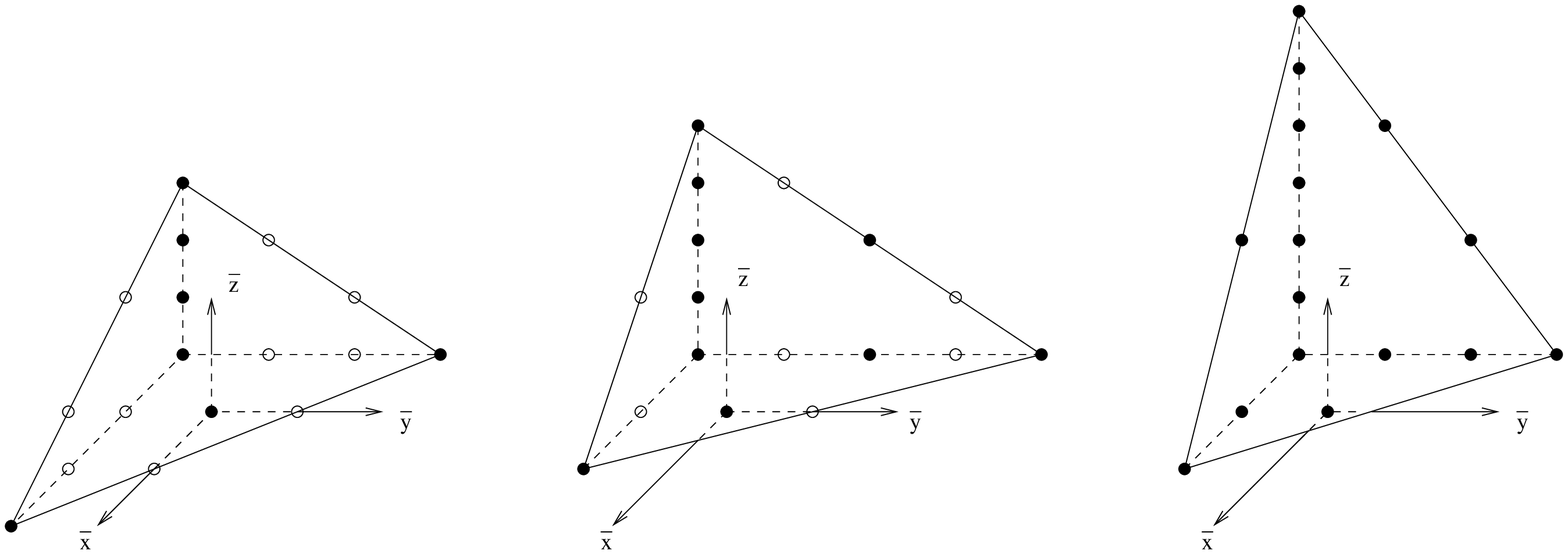}
\caption{Three maximal tops (the significance of $\bullet$ vs. $\circ$
will be explained in section 5)}\label{fig:e6e7e8}
\end{center}
\end{figure}

\ni
{\bf Case 3:} Neither of the above.\\
Then $b_0$ must be interior to a facet of $\top^*$ in such a way that
cases 0 and 2 do not apply.
In particular, after triangulating this facet $b_0$ must be interior to
one of the triangles $v_1\,v_2\,v_3$.
Applying $\p$, we see that $(0,0)$ is interior to $\D\subseteq F_0^*$
where $\D$ is the triangle with vertices $\p(v_i)$.
Every two dimensional lattice polytope whose only interior lattice
point is the origin is reflexive, so $\D$ must be one of the
triangles occurring in figure~\ref{fig:16pol} (numbers 1, 4, 10, 13, 16).
The linear relations among the vertices of these triangles imply
\bea v_1+v_2+v_3=3b_0&& \hbox{ for the triangles 1, 16},\\
   v_1+v_2+2v_3=4b_0&& \hbox{ for the triangles 4, 13},\\
   v_1+2v_2+3v_3=6b_0&&\hbox{ for triangle 10},
\eea
so $-1/z_0$ must divide one of the numbers $3,4,6$.
We can dismiss the following possibilities.\\
$z_0=-1$ implies case 0 by the lemma,\\
$z_0=-1/2$ is possible for triangles 4, 10, 13, but easily
seen to correspond to case 2,\\
$z_0=-1/3$ for triangles 10 or 16 can be reduced to triangle 1,\\
$z_0=-1/4$ for triangle 13 can be reduced to triangle 4.\\[2mm]
This leaves us with\\
{\bf a)} $z_0=-1/3$ for triangle 1,\\
{\bf b)} $z_0=-1/4$ for triangle 4,\\
{\bf c)} $z_0=-1/6$ for triangle 10.\\[2mm]
In each of these cases $\p(v_2)$ and $\p(v_3)$
generate the two dimensional lattice, so we can use (\ref{actg0}) to
put $v_2$ and $v_3$ at $z=0$ which forces $v_1$ to be at $z=-1$.
Let us denote by $P$ the prism over $\D$ cut off at the
$v_1v_2v_3$--plane. 
Then $P\subseteq \top^*$ implies $\top\subseteq P^*$, but $P^*$ is a
top with a finite number of lattice points. 
In other words, every top containing points at $\bar z >2$ is
contained in one of the three tops shown in figure~\ref{fig:e6e7e8}.
This implies that once a choice of $F_0^*$ as one of the 16 polygons
and (if possible) of $\D\subseteq F_0^*$ has been made, there is
only a finite number of consistent possibilities of assigning $\zm$ to
the remaining lattice points in $F_0^*$. 

The classification of all possible tops is then straightforward.
All that has to be done is to examine each of the 16 polygons of
figure~\ref{fig:16pol} with respect to all possibilities of applying
one of the cases 0, 1, 2, 3, taking care to avoid overcounting
wherever there are non-trivial automorphisms of $F_0^*$.
A complete list is given in the appendix.

\section{Discussion}

The results of the classification of course confirm the predictions made 
in section 3 on the structure of the edge diagram of a top at $\bar z \ge 1$.
By combining the arguments of the previous sections it is clear that
case 1 of our classification corresponds to $A_n^{(1)}$ algebras,
with 
\beq n+1=\sum_i l_i (\zm(p_i)+1) \end{equation} 
in the notation used around (\ref{latdist}).
Case 2a leads to $C_n^{(1)}$ diagrams where $n$ is just the parameter $a$ 
used there, 
and case 2b to $D_n^{(1)}$ and its folded versions where $n-4$ is 
the length of the edge at $\bar z = 2$, i.e. $2a$ or $2a+1$.
Cases 3a,b,c correspond to $E_6^{(1)}$, $E_7^{(1)}$, $E_8^{(1)}$
and their folded versions, respectively.

There is, however, a great difference between the occurrences of
untwisted and twisted Kac-Moody algebras as edge diagrams.
While each of the untwisted algebras occurs quite a number of times
and five of the reflexive polygons ($F_0^*$ one of 10, 11, 13, 14, 16
of figure~\ref{fig:16pol}) feature every possible untwisted algebra,
twisted algebras are quite rare.
Each of the diagrams of $E_6^{(2)}$, $A_2^{(2)}$ and $D_4^{(3)}$
occurs only once, and the members of the $D^{(2)}$ series occur twice.

There are only three pairs $(F_0,F_0^*)$ leading to diagrams of
twisted algebras, namely $(1,16)$, $(2,15)$ and $(4,13)$.
These reflexive pairs of polygons are quite special in several ways.
They are the only dual pairs where $F_0^*$ has no edge of length one;
this implies that no vertex of $F_0$ corresponds to a single section.
Moreover each of theses polygons is reflexive on two distinct
lattices, in such a way that the dual is the same polygon on the other
lattice. 
More precisely, polygon 1 is the same as polygon 16 on a sublattice of
index 3, and polygons 2 and 4 are the same as their respective duals on
sublattices of index 2.
We find that whenever twisted algebras occur, the corresponding tops
can be understood as coming from a restriction to a sublattice,
suitably extended in the third dimension.

Consider again the first top in figure~\ref{fig:e6e7e8}.
On the full lattice (with points both of the type $\circ$ and
$\bullet$) the edge diagram is an $E_6^{(1)}$ Dynkin diagram, but if
we consider the sublattice of index 3 determined only by points shown
as $\bullet$, we find the diagram of $D_4^{(3)}$.
This is the only twisted algebra coming from the pair $(1,16)$.

\begin{figure}[h]
\begin{center}
\includegraphics[width=12cm]{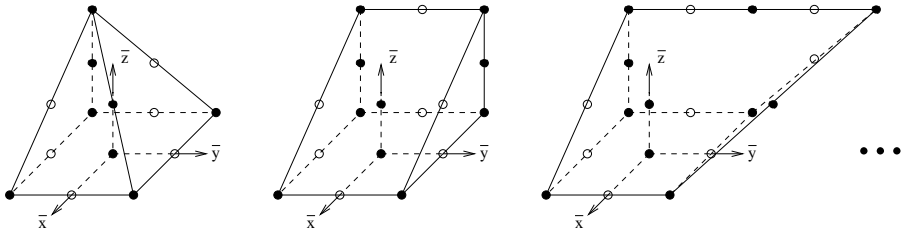}
\caption{A family of tops over squares}
\label{fig:square}
\end{center}
\end{figure}

The pair $(2,15)$ of reflexive squares gives rise to a family of
$D^{(2)}$ algebras as shown in figure~\ref{fig:square}.
Note, however, that the twisted diagrams $D_{i+3}^{(2)}$ with
$i=0,1,2,\ldots$ come from sublattice versions of tops corresponding
to $D_{2i+4}^{(1)}$, i.e. not the diagrams whose foldings produce the
twisted ones.

The situation is most intricate for the pair $(4,13)$ of reflexive
triangles.
In the second picture of figure~\ref{fig:e6e7e8}, passing
to the index 2 sublattice indicated by $\bullet$ means that we get an
$E_6^{(2)}$ diagram from an $E_7^{(1)}$ diagram.
\begin{figure}[h]
\begin{center}
\includegraphics[width=12.5cm]{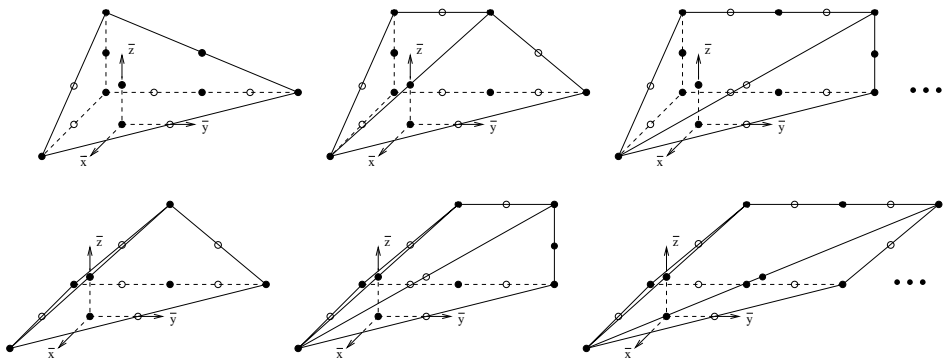}
\caption{Two families of tops over the dual pair $(4,13)$}
\label{fig:p112}
\end{center}
\end{figure}
Now consider the tops shown in figure~\ref{fig:p112}.
In the finer lattice including the points $\circ$, each member of the
first family shown in the upper row is isomorphic to the member of the
second family directly below; 
the transformation is (\ref{actg0}) with $a=0$, $b=1$, resulting in
tilting the top along the $y$ direction.
The corresponding algebras are $B_3^{(1)}$ and $D_{2i+4}^{(1)}$ with
$i\in\{1,2,\ldots\}$. 
In the coarser $\bullet$ lattice, however, this isomorphism is lost
as it would correspond to half integer parameters in (\ref{actg0}).
The first row gives a family of twisted diagrams $D_{i+3}^{(2)}$ with
$i\in\{0,1,2,\ldots\}$, as in the case of the square.
The second row gives another $D_{i+3}^{(2)}$ family for $i\ge 1$ with
a special case for $i=0$ where we have an $A_2^{(2)}$ diagram.
As an additional subtlety, the tops for odd $i$ are nevertheless
isomorphic, namely by an isomorphism that acts on $F_0$ by
swapping the two vertices of the long edge.
For this reason we have listed these tops in table~\ref{tab:results13}
in the appendix as two families of type $D_{2i+3}^{(2)}$ and one
family of type $D_{2i+4}^{(2)}$.

The fact that the tops giving rise to twisted algebras all correspond 
to pairs of lattices has the following interpretation.
Passing from a coarser lattice $N$ to a finer lattice $N'$ in a way that 
is compatible with the structure of the fan means passing from a toric 
variety to a quotient of this variety by a finite abelian group isomorphic 
to $N'/N$.
In the present cases this group is just $\IZ_3$ or $\IZ_2$.
So the varieties corresponding to diagrams related to twisted algebras
allow group actions in such a way that taking the quotient would 
take us to the variety defined by the finer lattice.

Finally let us discuss the relation between tops and string theory.
The simplest case is that of taking an elliptic K3 surface whose toric 
polytope consists of two tops.
Blowing down all divisors except two at $\bar z = \pm 1$ will result in
the occurrence of two ADE singularities (at $W=0/\infty$) corresponding 
to the unfolded diagram.
Compactifying M-theory on such a space leads to a theory where the 
generic abelian gauge group is enhanced in such a way that the corresponding 
ADE groups arise.
The same groups also arise in compactifications of type IIA and F-theory 
without background fields.

If a top is part of a diagram describing an elliptically fibered Calabi--Yau
threefold or fourfold, the generic local geometry is a product of a 
neighborhood $U$ in $\IC$ or $\IC^2$ with the two dimensional geometry 
featuring the ADE pattern;
every exceptional rational curve gives rise to a divisor $U\times \IP^1$.
Globally several of these divisors may correspond to a single irreducible 
divisor (this is the non-split case of \cite{BIKMSV}).
Clearly this can happen only if the different $\IP^1$'s all come from the 
same lattice point in the toric diagram.
In this case there are special loci in the base space over which some or all 
of the $\IP^1$'s coincide. 
Monodromy around these loci will interchange the $\IP^1$'s.
Upon compactifications of the same theories as above this results in non 
simply laced gauge groups \cite{AG}, again of the type determined by the toric 
diagram.

These constructions are conjectured to be dual to compactifications 
of heterotic strings;
in particular the K3 compactifications are dual to toroidal heterotic 
compactifications with maximal rank of the gauge group.
However, the heterotic moduli space also contains components of reduced 
rank of the gauge theory \cite{CHL,CP}.
These require non-standard IIA and F-theory compactifications as duals.
In the IIA case these are compactifications on orbifolds in the presence 
of a non-trivial RR background \cite{SS}; 
this has an M-theory lift where the orbifolding also acts by a shift on the 
$S^1$ in the eleventh dimension.
In terms of F-theory a non-generic monodromy group together with non-trivial
$B_{\m\n}$ flux in the underlying IIB theory is required \cite{BPS};
F-theory duals were also considered in \cite{Pa,BKMT}.
We will now see how these requirements are met by the tops that give rise 
to twisted Kac-Moody algebras; that these algebras should play a role
on the heterotic side was predicted in \cite{LMST}.

Consider once more the second diagram of figure~\ref{fig:e6e7e8}.
To be specific let us assume that it is part of the reflexive polytope that 
is obtained by adding the `bottom' that is the reflection of the top through 
the $xy$-plane (this reflexive polytope was also considered in \cite{BKMT},
but our discussion will be different).
Passing from the space determined by the $\bullet$ diagram to its $\IZ_2$ 
orbifold means that we get singularities that may be resolved by blowing
up along the divisors indicated by $\circ$.
In a standard IIA compactification passing to the orbifold means that one 
loses as many non-algebraic cycles as there are $\circ$ cycles and the rank
of the gauge group remains 24.
In the compactification with RR background the $\circ$ cycles do not 
contribute and we end up with reduced rank.
By counting with the right multiplicities (one for every $\circ$ point 
in an edge except $(0,1,0)$ which has multiplicity two) we get a rank 
reduction of eight as it should be \cite{CHL,CL}.
The same reflexive polytope allows for a second fibration structure
with the fiber determined by the polygon at $x+y=0$.
Now the orbifolding acts by changing the $B_3^{(1)}$ top at $x+y>0$ to
a trivial top and the $D_{10}^{(1)}$ top at $x+y<0$ to a $C_6^{(1)}$ top.
This is the toric description of the involution discussed in section 2 of 
\cite{BPS}.

In a similar manner the diagrams of figures \ref{fig:square} and 
\ref{fig:p112} can be used to construct theories with rank reduction of 
eight.
Taking the first top in figure~\ref{fig:e6e7e8} together with its mirror 
image, we get a K3 with orbifold group $\IZ_3$. 
According to \cite{CL} this should lead to a rank reduction of 12;
with the same counting as before this is indeed confirmed.

The gauge groups that we get are again non simply laced with a mechanism 
very similar to the one we encountered before.
In the M-theory picture we compactify on $(K3\times S^1)/G$ which is a smooth 
non-trivial bundle over $S^1$ with fiber the original $K3$.
So locally over a neighborhood in $S^1$ we get all the $\IP^1$'s of the 
untwisted diagram but upon going around the $S^1$ they are permuted.
For obtaining gauge groups we blow down the toric divisors corresponding
to all points of a top except for one at $\bar z =1$.
The collapsed cycles belong to ordinary ADE Dynkin diagrams (the heights
play no role here) that are folded by the permutations.
In this way $E_6^{(2)}$, $D_{n+1}^{(2)}$, $A_2^{(2)}$ and $D_4^{(3)}$
give rise to the groups $F_4$, $B_n$, $A_1$ and $G_2$, respectively.

Tops from the families of figures \ref{fig:square} and \ref{fig:p112} that do 
not fit into
three dimensional reflexive polytopes should play a role in theories dual to 
non-toroidal compactifications of CHL strings \cite{KKO}.

\section*{Appendix: Classification results}
The following tables provide the complete list of possible tops $\top$,
in terms of their duals $\top^*$.
The first column of each table identifies the polygon $F_0^*$ to
which $\top^*$ projects according to the numbering in figure~\ref{fig:16pol}.
The following columns give $z_k:=\zm(b_k)$, where $b_0$ is the
origin of $F_0^*$ and the other $b_k$ are the lattice points at the
boundary, starting at the `12 o'clock position' and proceeding clockwise.
The parameter $i$ takes values in $\{0,1,\ldots\}$.
For the elements of the $A$-series (last row for each choice of
$F_0^*$) the $z_k$ are assumed to satisfy the inequalities determined
by (\ref{latdist}).

The last column indicates the affine Kac-Moody algebra to which $\top$
corresponds, with the superscript ${}^{(1)}$ suppressed for the untwisted
cases.
The case $i=0$ can be special in the sense that the corresponding
Dynkin diagram does not belong to the general family (this occurs when
edges that are distinct for $i>0$ merge when $i=0$), or that
$\top$ is the same for some other family.
In either of these cases we also display the $i=0$ algebra, with
superscripts ${}^a$ or ${}^b$ for repeated non-trivial cases.

Any two duals of tops in our tables are different, with the
following exceptions.
The trivial top (case 0) may occur more than once as the $i=0$ case of
a $C_i$ series; $C_0$ always means the trivial case and its repetitions
are not separately indicated.
If a non-trivial top occurs more than once as an $i=0$ case, it gets a
superscript ${}^a$ or ${}^b$ which is the same for each occurrence.
For the $A$-cases, we have not eliminated the equivalences coming from
automorphisms of $F_0^*$.

\newpage
${}$
\vfill

\begin{table}[!b]
\begin{center}
\begin{footnotesize}
\begin{tabular}{|c|cccccc|c|}\hline
%$F_0^*$&\multicolumn{6}{c}{Minimal points}\vline&Lie\\[0cm]
%\cline{2-7}
$F_0^*$&$z_0$&$z_1$&$z_2$&$z_3$&$z_4$&$z_5$&AKMA\\[0cm]
\hline\hline

1&$-1/3$&$-1$&$-1$&$1$&-&-&$E\sst 6$\\[0cm]
\cline{2-8}
&$-1$&$-1$&$-1$&$-1$&-&-&$C\sst {0}$\\[0cm]
\cline{2-8}
&$-1$&$-1$&$-1$&$z_3$&-&-&$A\sst {3z_3+2}$\\[0cm]
\hline\hline

2&$-1/2$&$-1$&$-1$&$0$&$i$&-&$D\sst {2i+4}$\\[0cm]
\cline{2-8}
&$-1$&$-1$&$-1$&$-1$&$i-1$&-&$C\sst i$\\[0cm]
\cline{2-8}
&$-1$&$-1$&$-1$&$z_3$&$z_4$&-&$A\sst {2z_3+2z_4+3}$\\[0cm]
\hline\hline

3&$-1/3$&$-1$&$-1$&$1$&$1$&-&$E\sst 6$\\[0cm]
\cline{2-8}
&$-1/2$&$-1$&$-1$&$0$&$i+1$&-&$D\sst {2i+5}$\\[0cm]
\cline{2-8}
&$-1$&$-1$&$-1$&$-1$&$i-1$&-&$C\sst i$\\[0cm]
\cline{2-8}
&$-1$&$-1$&$-1$&$z_3$&$z_4$&-&$A\sst {z_3+2z_4+2}$\\[0cm]
\hline\hline

4&$-1/4$&$-1$&$-1$&$1/2$&$2$&-&$E\sst 7$\\[0cm]
\cline{2-8}
&$-1/2$&$-1$&$-1$&$0$&$i+1$&-&$D\sst {2i+4} / B\sst 3$\\[0cm]
\cline{2-8}
&$-1$&$-1$&$-1$&$-1$&$i-1$&-&$C\sst i$\\[0cm]
\cline{2-8}
&$-1$&$-1$&$-1$&$z_3$&$z_4$&-&$A\sst {2z_4+1}$\\[0cm]
\hline\hline

5&$-1/3$&$-1$&$-1$&$1$&$1$&$-1$&$E\sst 6$\\[0cm]
\cline{2-8}
&$-1/2$&$-1$&$-1$&$0$&$i$&$i$&$B\sst {2i+4}$\\[0cm]
\cline{2-8}
&$-1/2$&$-1$&$-1$&$0$&$i + 1$&$i$&$B\sst {2i+5}$\\[0cm]
\cline{2-8}
&$-1$&$-1$&$-1$&$-1$&$i-1$&$i-1$&$C\sst i$\\[0cm]
\cline{2-8}
&$-1$&$-1$&$-1$&$z_3$&$z_4$&$z_5$&$A\sst {2z_3+z_4+z_5+3}$\\[0cm]

\hline\hline

6&$-1/4$&$-1$&$-1$&$1/2$&$2$&$2$&$E\sst 7$\\[0cm]
\cline{2-8}
&$-1/3$&$-1$&$-1$&$1/2$&$2$&$1$&$E\sst 6$\\[0cm]
\cline{2-8}
&$-1/3$&$-1$&$-1$&$1$&$3$&$1$&$E\sst 6$\\[0cm]
\cline{2-8}
&$-1/2$&$-1$&$-1$&$0$&$i+1$&$i+1$&$D\sst {2i+4} / B\sst {3}$\\[0cm]
\cline{2-8}
&$-1/2$&$-1$&$-1$&$0$&$i+2$&$i+1$&$D\sst {2i+5}$\\[0cm]
\cline{2-8}
&$-1$&$-1$&$-1$&$-1$&$i-1$&$i-1$&$C\sst i$\\[0cm]
\cline{2-8}
&$-1$&$-1$&$-1$&$z_3$&$z_4$&$z_5$&$A\sst {z_4+z_5+1}$\\[0cm]
\hline\hline

\end{tabular}
\caption{Duals $\top^*$ of tops with $F_0^*$ one of the polygons
$1,\ldots,6$ of figure~\ref{fig:16pol}.}
\label{tab:results1to6}
\end{footnotesize}
\end{center}
\end{table}

%\vfill
%${}$

\newpage
${}$
\vfill

\begin{table}[!b]
\begin{center}
\begin{footnotesize}
\begin{tabular}{|c|ccccccc|c|}
\hline

$F_0^*$&$z_0$&$z_1$&$z_2$&$z_3$&$z_4$&$z_5$&$z_6$&AKMA\\[-0mm]
\hline\hline

7&$-1/3$&$-1$&$-1$&$1$&$1$&$1$&$-1$&$E\sst 6$\\[-0mm]
\cline{2-9}
&$-1/2$&$-1$&$-1$&$0$&$0$&$i$&$i$&$D\sst {2i+4}$\\[-0mm]
\cline{2-9}
&$-1/2$&$-1$&$-1$&$0$&$0$&$i+1$&$i$&$D\sst {2i+5}$\\[-0mm]
\cline{2-9}
&$-1$&$-1$&$-1$&$-1$&$-1$&$i-1$&$i-1$&$C\sst i$\\[-0mm]
\cline{2-9}

&$-1$&$-1$&$-1$&$z_3$&$z_4$&$z_5$&$z_6$&$A\sst {z_3+z_4+z_5+z_6+3}$\\[-0mm]
\hline\hline

8&$-1/4$&$-1$&$-1$&$1/2$&$2$&$3/2$&$1$&$E\sst 7$\\[-0mm]
\cline{2-9}
&$-1/4$&$-1$&$-1$&$1/2$&$2$&$2$&$2$&$E\sst 7$\\[-0mm]
\cline{2-9}
&$-1/3$&$-1$&$-1$&$1/2$&$2$&$1$&$1$&$E\sst 6$\\[-0mm]
\cline{2-9}
&$-1/3$&$-1$&$-1$&$1$&$3$&$1$&$1$&$E\sst 6$\\[-0mm]
\cline{2-9}

&$-1/2$&$-1$&$-1$&$0$&$i+1$&$i+1/2$&$i$&$D\sst {2i+4} \yy B\sst {3}^a\xx$\\[-0mm]
\cline{2-9}
&$-1/2$&$-1$&$-1$&$0$&$i+2$&$i+1$&$i$&$D\sst {2i+4} \yy D\sst 4^a\xx$\\[-0mm]
\cline{2-9}
&$-1/2$&$-1$&$-1$&$0$&$i+1$&$i+1$&$i+1$&$B\sst {2i+4} \yy B\sst 3\xx$\\[-0mm]
\cline{2-9}
&$-1/2$&$-1$&$-1$&$i$&$2i+2$&$i+1$&$0$&$B\sst {2i+4} \yy D\sst 4^a\xx$\\[-0mm]
\cline{2-9}
&$-1/2$&$-1$&$-1$&$0$&$i+2$&$i+1$&$i+1$&$B\sst {2i+5}$\\[-0mm]
\cline{2-9}
&$-1/2$&$-1$&$-1$&$i+1$&$2i+3$&$i+1$&$0$&$B\sst {2i+5}$\\[-0mm]
\cline{2-9}
&$-1/2$&$-1$&$-1$&$i/2$&$i+1$&$(i+1)/2$&$0$&$B\sst {i+3} \yy B\sst {3}^a\xx$\\[-0mm]
\cline{2-9}
&$-1$&$-1$&$-1$&$-1$&$i-1$&$i-1$&$i-1$&$C\sst i$\\[-0mm]
\cline{2-9}
&$-1$&$-1$&$-1$&$i-1$&$2i-1$&$i-1$&$-1$&$C\sst i$\\[-0mm]
\cline{2-9}
&$-1$&$-1$&$-1$&$z_3$&$z_4$&$z_5$&$z_6$&$A\sst {z_4+z_6+1}$\\[-0mm]
\hline\hline

9&$-1/4$&$2$&$-1$&$-1$&$-1$&$1/2$&$2$&$E\sst 7$\\[-0mm]
\cline{2-9}
&$-1/3$&$-1$&$-1$&$1$&$1$&$1/2$&$0$&$E\sst 6$\\[-0mm]
\cline{2-9}
&$-1/3$&$-1$&$-1$&$1$&$1$&$1$&$1$&$E\sst 6$\\[-0mm]
\cline{2-9}

&$-1/2$&$-1$&$-1$&$0$&$i + 1$&$i$&$i$&$D\sst {2i+4} \yy D\sst 4^a\xx$\\[-0mm]
\cline{2-9}
&$-1/2$&$-1$&$-1$&$i$&$i + 1$&$0$&$0$&$D\sst {2i+4} \yy D\sst 4^a\xx$\\[-0mm]
\cline{2-9}
&$-1/2$&$-1$&$-1$&$i+1$&$i+1$&$0$&$-1$&$D\sst {2i+4} \yy B\sst 3\xx$\\[-0mm]
\cline{2-9}
&$-1/2$&$-1$&$-1$&$0$&$i+1$&$i+1/2$&$i$&$D\sst {2i+5}$\\[-0mm]
\cline{2-9}
&$-1/2$&$-1$&$-1$&$0$&$i+1$&$i+1$&$i+1$&$D\sst {2i+5}$\\[-0mm]
\cline{2-9}
&$-1/2$&$-1$&$-1$&$i+1$&$i+1$&$0$&$0$&$D\sst {2i+5}$\\[-0mm]
\cline{2-9}

&$-1$&$-1$&$-1$&$-1$&$i-1$&$i-1$&$i-1$&$C\sst i$\\[-0mm]
\cline{2-9}
&$-1$&$-1$&$-1$&$i-1$&$i-1$&$-1$&$-1$&$C\sst i$\\[-0mm]
\cline{2-9}
&$-1$&$-1$&$-1$&$z_3$&$z_4$&$z_5$&$z_6$&$A\sst {z_3+z_4+z_6+2}$\\[-0mm]
\hline\hline

10&$-1/6$&$-1$&$-1$&$2/3$&$7/3$&$4$&$3/2$&$E\sst 8$\\[-0mm]
\cline{2-9}
&$-1/4$&$-1$&$-1$&$1/2$&$2$&$4$&$3/2$&$E\sst 7$\\[-0mm]
\cline{2-9}
&$-1/4$&$-1$&$-1$&$1/2$&$2$&$5$&$2$&$E\sst 7$\\[-0mm]
\cline{2-9}
&$-1/3$&$-1$&$-1$&$1/3$&$5/3$&$3$&$1$&$F\sst 4$\\[-0mm]
\cline{2-9}
&$-1/3$&$-1$&$-1$&$1/2$&$2$&$4$&$1$&$E\sst 6$\\[-0mm]
\cline{2-9}
&$-1/3$&$-1$&$-1$&$1$&$3$&$5$&$1$&$E\sst 6$\\[-0mm]
\cline{2-9}

&$-1/2$&$-1$&$-1$&$0$&$i+1$&$2i+3$&$i+1$&$D\sst {2i+4} \yy B\sst 3\xx$\\[-0mm]
\cline{2-9}
&$-1/2$&$-1$&$-1$&$0$&$i +2$&$2i+4$&$i+1$&$D\sst {2i+5}$\\[-0mm]
\cline{2-9}
&$-1/2$&$-1$&$-1$&$0$&$i/2+1$&$i+2$&$(i+1)/2$&$B\sst {i+3} \yy G\sst 2\xx$\\[-0mm]
\cline{2-9}

&$-1$&$-1$&$-1$&$-1$&$i-1$&$2i-1$&$i-1$&$C\sst i$\\[-0mm]
\cline{2-9}

&$-1$&$-1$&$-1$&$z_3$&$z_4$&$z_5$&$z_6$&$A\sst {z_5}$\\[-0mm]
\hline\hline

\end{tabular}
\caption{Duals $\top^*$ of tops with $F_0^*$ one of the polygons
$7,8,9,10$ of figure~\ref{fig:16pol}.}\label{tab:results7to10}
\end{footnotesize}
\end{center}
\end{table}

\newpage

%\vspace*{\fill}

\begin{table}[!b]
\begin{center}
\begin{footnotesize}
\begin{tabular}{|c|cccccccc|c|}
\hline

\!$F_0^*$\!\!&\!\!$z_0$\!\!&\!\!$z_1$\!\!&\!\!$z_2$\!\!&\!\!$z_3$\!\!&\!\!$z_4$\!\!&\!\!$z_5$\!\!&\!\!$z_6$\!\!&\!\!$z_7$\!&\!AKMA\!\\[-.3mm]
\hline\hline

\!11\!&\!$-1/6$\!\!&\!\!$-1$\!\!&\!\!$-1$\!\!&\!\!$-1$\!\!&\!\!$2/3$\!\!&\!\!$7/3$\!\!&\!\!$4$\!\!&\!\!$3/2$\!\!&\!\!$E\sst 8$\!\\[-.3mm]
\cline{2-10}
\!&\!$-1/4$\!\!&\!\!$-1$\!\!&\!\!$-1$\!\!&\!\!$-1$\!\!&\!\!$1/2$\!\!&\!\!$2$\!\!&\!\!$4$\!\!&\!\!$3/2$\!\!&\!\!$E\sst 7$\!\\[-.3mm]
\cline{2-10}
\!&\!$-1/4$\!\!&\!\!$-1$\!\!&\!\!$-1$\!\!&\!\!$-1$\!\!&\!\!$1/2$\!\!&\!\!$2$\!\!&\!\!$5$\!\!&\!\!$2$\!\!&\!\!$E\sst 7$\!\\[-.3mm]
\cline{2-10}
\!&\!$-1/4$\!\!&\!\!$-1$\!\!&\!\!$-1$\!\!&\!\!$0$\!\!&\!\!$2/3$\!\!&\!\!$4/3$\!\!&\!\!$2$\!\!&\!\!$1/2$\!\!&\!\!$E\sst 7$\!\\[-.3mm]
\cline{2-10}
\!&\!$-1/4$\!\!&\!\!$-1$\!\!&\!\!$-1$\!\!&\!\!$1$\!\!&\!\!$1$\!\!&\!\!$3/2$\!\!&\!\!$2$\!\!&\!\!$1/2$\!\!&\!\!$E\sst 7$\!\\[-.3mm]
\cline{2-10}
\!&\!$-1/4$\!\!&\!\!$-1$\!\!&\!\!$-1$\!\!&\!\!$2$\!\!&\!\!$2$\!\!&\!\!$2$\!\!&\!\!$2$\!\!&\!\!$1/2$\!\!&\!\!$E\sst 7$\!\\[-.3mm]
\cline{2-10}
\!&\!$-1/3$\!\!&\!\!$-1$\!\!&\!\!$-1$\!\!&\!\!$-1$\!\!&\!\!$1/3$\!\!&\!\!$5/3$\!\!&\!\!$3$\!\!&\!\!$1$\!\!&\!\!$F\sst 4$\!\\[-.3mm]
\cline{2-10}
\!&\!$-1/3$\!\!&\!\!$-1$\!\!&\!\!$1$\!\!&\!\!$1$\!\!&\!\!$1/2$\!\!&\!\!$0$\!\!&\!\!$0$\!\!&\!\!$-1$\!\!&\!\!$E\sst 6$\!\\[-.3mm]
\cline{2-10}
\!&\!$-1/3$\!\!&\!\!$-1$\!\!&\!\!$1$\!\!&\!\!$1$\!\!&\!\!$1$\!\!&\!\!$1$\!\!&\!\!$1$\!\!&\!\!$-1$\!\!&\!\!$E\sst 6$\!\\[-.3mm]
\cline{2-10}
\!&\!$-1/3$\!\!&\!\!$-1$\!\!&\!\!$-1$\!\!&\!\!$0$\!\!&\!\!$1/2$\!\!&\!\!$1$\!\!&\!\!$2$\!\!&\!\!$1/2$\!\!&\!\!$E\sst 6$\!\\[-.3mm]
\cline{2-10}
\!&\!$-1/3$\!\!&\!\!$-1$\!\!&\!\!$-1$\!\!&\!\!$1$\!\!&\!\!$1$\!\!&\!\!$1$\!\!&\!\!$2$\!\!&\!\!$1/2$\!\!&\!\!$E\sst 6$\!\\[-.3mm]
\cline{2-10}
\!&\!$-1/3$\!\!&\!\!$-1$\!\!&\!\!$-1$\!\!&\!\!$0$\!\!&\!\!$1/2$\!\!&\!\!$1$\!\!&\!\!$3$\!\!&\!\!$1$\!\!&\!\!$E\sst 6$\!\\[-.3mm]
\cline{2-10}
\!&\!$-1/3$\!\!&\!\!$-1$\!\!&\!\!$-1$\!\!&\!\!$1$\!\!&\!\!$1$\!\!&\!\!$1$\!\!&\!\!$3$\!\!&\!\!$1$\!\!&\!\!$E\sst 6$\!\\[-.3mm]
\cline{2-10}

\!&\!$-1/2$\!\!&\!\!$-1$\!\!&\!\!$-1$\!\!&\!\!$0$\!\!&\!\!$0$\!\!&\!\!$i+1$\!\!&\!\!$2i+2$\!\!&\!\!$i$\!\!&\!\!$D\sst {2i+4} \yy D\sst 4^a\xx$\!\\[-.3mm]
\cline{2-10}
\!&\!$-1/2$\!\!&\!\!$-1$\!\!&\!\!$-1$\!\!&\!\!$-1$\!\!&\!\!$0$\!\!&\!\!$i+1$\!\!&\!\!$2i+3$\!\!&\!\!$i+1$\!\!&\!\!$D\sst {2i+4} \yy B\sst 3\xx$\!\\[-.3mm]
\cline{2-10}
\!&\!$-1/2$\!\!&\!\!$-1$\!\!&\!\!$-1$\!\!&\!\!$i$\!\!&\!\!$i$\!\!&\!\!$i+1/2$\!\!&\!\!$i+1$\!\!&\!\!$0$\!\!&\!\!$D\sst {2i+4} \yy B\sst 3^a\xx$\!\\[-.3mm]
\cline{2-10}
\!&\!$-1/2$\!\!&\!\!$-1$\!\!&\!\!$-1$\!\!&\!\!$i$\!\!&\!\!$i$\!\!&\!\!$i+1$\!\!&\!\!$i+2$\!\!&\!\!$0$\!\!&\!\!$D\sst {2i+4} \yy D\sst 4^a\xx$\!\\[-.3mm]
\cline{2-10}
\!&\!$-1/2$\!\!&\!\!$-1$\!\!&\!\!$-1$\!\!&\!\!$i+1$\!\!&\!\!$i+1$\!\!&\!\!$i+1$\!\!&\!\!$i+1$\!\!&\!\!$0$\!\!&\!\!$D\sst {2i+4} \yy B\sst 3\xx$\!\\[-.3mm]
\cline{2-10}

\!&\!$-1/2$\!\!&\!\!$-1$\!\!&\!\!$-1$\!\!&\!\!$0$\!\!&\!\!$0$\!\!&\!\!$i+1$\!\!&\!\!$2i+3$\!\!&\!\!$i+1$\!\!&\!\!$D\sst {2i+5}$\!\\[-.3mm]
\cline{2-10}
\!&\!$-1/2$\!\!&\!\!$-1$\!\!&\!\!$-1$\!\!&\!\!$-1$\!\!&\!\!$0$\!\!&\!\!$i +2$\!\!&\!\!$2i+4$\!\!&\!\!$i+1$\!\!&\!\!$D\sst {2i+5}$\!\\[-.3mm]
\cline{2-10}
\!&\!$-1/2$\!\!&\!\!$-1$\!\!&\!\!$-1$\!\!&\!\!$i$\!\!&\!\!$i+1/2$\!\!&\!\!$i+1$\!\!&\!\!$i+2$\!\!&\!\!$0$\!\!&\!\!$D\sst {2i+5}$\!\\[-.3mm]
\cline{2-10}
\!&\!$-1/2$\!\!&\!\!$-1$\!\!&\!\!$-1$\!\!&\!\!$i$\!\!&\!\!$i+1$\!\!&\!\!$i+2$\!\!&\!\!$i+3$\!\!&\!\!$0$\!\!&\!\!$D\sst {2i+5}$\!\\[-.3mm]
\cline{2-10}
\!&\!$-1/2$\!\!&\!\!$-1$\!\!&\!\!$-1$\!\!&\!\!$i+1$\!\!&\!\!$i+1$\!\!&\!\!$i+1$\!\!&\!\!$i+2$\!\!&\!\!$0$\!\!&\!\!$D\sst {2i+5}$\!\\[-.3mm]
\cline{2-10}

\!&\!$-1/2$\!\!&\!\!$-1$\!\!&\!\!$-1$\!\!&\!\!$0$\!\!&\!\!$0$\!\!&\!\!$(i+1)/2$\!\!&\!\!$i+1$\!\!&\!\!$i/2$\!\!&\!\!$B\sst {i+3} \yy B\sst 3^a\xx$\!\\[-.3mm]
\cline{2-10}
\!&\!$-1/2$\!\!&\!\!$-1$\!\!&\!\!$-1$\!\!&\!\!$-1$\!\!&\!\!$0$\!\!&\!\!$i/2+1$\!\!&\!\!$i+2$\!\!&\!\!$(i+1)/2$\!\!&\!\!$B\sst {i+3} \yy G\sst 2\xx$\!\\[-.3mm]
\cline{2-10}

\!&\!$-1$\!\!&\!\!$-1$\!\!&\!\!$-1$\!\!&\!\!$i-1$\!\!&\!\!$i-1$\!\!&\!\!$i-1$\!\!&\!\!$i-1$\!\!&\!\!$-1$\!\!&\!\!$C\sst i$\!\\[-.3mm]
\cline{2-10}
\!&\!$-1$\!\!&\!\!$-1$\!\!&\!\!$-1$\!\!&\!\!$-1$\!\!&\!\!$-1$\!\!&\!\!$i-1$\!\!&\!\!$2i-1$\!\!&\!\!$i-1$\!\!&\!\!$C\sst i$\!\\[-.3mm]
\cline{2-10}

\!&\!$-1$\!\!&\!\!$-1$\!\!&\!\!$-1$\!\!&\!\!$z_3$\!\!&\!\!$z_4$\!\!&\!\!$z_5$\!\!&\!\!$z_6$\!\!&\!\!$z_7$\!\!&\!\!$A\sst {z_3+z_6+1}$\!\\[-.3mm]
\hline\hline

\!12\!\!&\!\!$-1/4$\!\!&\!\!$-3/2$\!\!&\!\!$-2$\!\!&\!\!$-1$\!\!&\!\!$2$\!\!&\!\!$2$\!\!&\!\!$1/2$\!\!&\!\!$-1$\!\!&\!\!$E\sst 7$\!\\[-.3mm]
\cline{2-10}
\!\!&\!\!$-1/4$\!\!&\!\!$-1$\!\!&\!\!$-1$\!\!&\!\!$-1$\!\!&\!\!$2$\!\!&\!\!$2$\!\!&\!\!$1/2$\!\!&\!\!$-1$\!\!&\!\!$E\sst 7$\!\\[-.3mm]
\cline{2-10}
\!\!&\!\!$-1/3$\!\!&\!\!$-3/2$\!\!&\!\!$-2$\!\!&\!\!$-1$\!\!&\!\!$1$\!\!&\!\!$2$\!\!&\!\!$1/2$\!\!&\!\!$-1$\!\!&\!\!$E\sst 6$\!\\[-.3mm]
\cline{2-10}
\!\!&\!\!$-1/3$\!\!&\!\!$-3/2$\!\!&\!\!$-2$\!\!&\!\!$-1$\!\!&\!\!$1$\!\!&\!\!$3$\!\!&\!\!$1$\!\!&\!\!$-1$\!\!&\!\!$E\sst 6$\!\\[-.3mm]
\cline{2-10}
\!\!&\!\!$-1/3$\!\!&\!\!$-1$\!\!&\!\!$-1$\!\!&\!\!$-1$\!\!&\!\!$1$\!\!&\!\!$3$\!\!&\!\!$1$\!\!&\!\!$-1$\!\!&\!\!$E\sst 6$\!\\[-.3mm]
\cline{2-10}
\!\!&\!\!$-1/3$\!\!&\!\!$-1$\!\!&\!\!$-1$\!\!&\!\!$-1$\!\!&\!\!$1$\!\!&\!\!$1$\!\!&\!\!$1/2$\!\!&\!\!$0$\!\!&\!\!$E\sst 6$\!\\[-.3mm]
\cline{2-10}
\!\!&\!\!$-1/3$\!\!&\!\!$-1$\!\!&\!\!$-1$\!\!&\!\!$-1$\!\!&\!\!$1$\!\!&\!\!$1$\!\!&\!\!$1$\!\!&\!\!$1$\!\!&\!\!$E\sst 6$\!\\[-.3mm]
\cline{2-10}

\!\!&\!\!$-1/2$\!\!&\!\!$-1$\!\!&\!\!$-1$\!\!&\!\!$0$\!\!&\!\!$0$\!\!&\!\!$i$\!\!&\!\!$i-1/2$\!\!&\!\!$i-1$\!\!&\!\!$D\sst {2i+4}\yy B\sst 3^a\xx$\!\\[-.3mm]
\cline{2-10}
\!\!&\!\!$-1/2$\!\!&\!\!$-1$\!\!&\!\!$-1$\!\!&\!\!$0$\!\!&\!\!$0$\!\!&\!\!$i$\!\!&\!\!$i$\!\!&\!\!$i$\!\!&\!\!$D\sst {2i+4}\yy D\sst 4^a\xx$\!\\[-.3mm]
\cline{2-10}
\!\!&\!\!$-1/2$\!\!&\!\!$-1$\!\!&\!\!$-1$\!\!&\!\!$0$\!\!&\!\!$0$\!\!&\!\!$i+1$\!\!&\!\!$i$\!\!&\!\!$i-1$\!\!&\!\!$D\sst {2i+4}\yy B\sst 3\xx$\!\\[-.3mm]
\cline{2-10}
\!\!&\!\!$-1/2$\!\!&\!\!$-1$\!\!&\!\!$-1$\!\!&\!\!$-1$\!\!&\!\!$0$\!\!&\!\!$i+1$\!\!&\!\!$i$\!\!&\!\!$i$\!\!&\!\!$D\sst {2i+4}\yy D\sst 4\xx$\!\\[-.3mm]
\cline{2-10}
\!\!&\!\!$-1/2$\!\!&\!\!$i$\!\!&\!\!$-1$\!\!&\!\!$-1$\!\!&\!\!$0$\!\!&\!\!$0$\!\!&\!\!$i$\!\!&\!\!$2i+1$\!\!&\!\!$D\sst {2i+4}\yy D\sst 4^a\xx$\!\\[-.3mm]
\cline{2-10}
\!\!&\!\!$-1/2$\!\!&\!\!$-1$\!\!&\!\!$-1$\!\!&\!\!$0$\!\!&\!\!$0$\!\!&\!\!$i+1$\!\!&\!\!$i$\!\!&\!\!$i$\!\!&\!\!$D\sst {2i+5}$\!\\[-.3mm]
\cline{2-10}
\!\!&\!\!$-1/2$\!\!&\!\!$-1$\!\!&\!\!$-1$\!\!&\!\!$-1$\!\!&\!\!$0$\!\!&\!\!$i+1$\!\!&\!\!$i+1/2$\!\!&\!\!$i$\!\!&\!\!$D\sst {2i+5}$\!\\[-.3mm]
\cline{2-10}
\!\!&\!\!$-1/2$\!\!&\!\!$-1$\!\!&\!\!$-1$\!\!&\!\!$-1$\!\!&\!\!$0$\!\!&\!\!$i+1$\!\!&\!\!$i+1$\!\!&\!\!$i+1$\!\!&\!\!$D\sst {2i+5}$\!\\[-.3mm]
\cline{2-10}
\!\!&\!\!$-1/2$\!\!&\!\!$-1$\!\!&\!\!$-1$\!\!&\!\!$-1$\!\!&\!\!$0$\!\!&\!\!$i+2$\!\!&\!\!$i+1$\!\!&\!\!$i$\!\!&\!\!$D\sst {2i+5}$\!\\[-.3mm]
\cline{2-10}
\!\!&\!\!$-1/2$\!\!&\!\!$i$\!\!&\!\!$-1$\!\!&\!\!$-1$\!\!&\!\!$0$\!\!&\!\!$0$\!\!&\!\!$i+1$\!\!&\!\!$2i+2$\!\!&\!\!$D\sst {2i+5}$\!\\[-.3mm]
\cline{2-10}
\!\!&\!\!$-1/2$\!\!&\!\!$(i-1)/2$\!\!&\!\!$-1$\!\!&\!\!$-1$\!\!&\!\!$0$\!\!&\!\!$0$\!\!&\!\!$i/2$\!\!&\!\!$i$\!\!&\!\!$B\sst {i+3}\yy B\sst 3^a\xx$\!\\[-.3mm]
\cline{2-10}

\!\!&\!\!$-1$\!\!&\!\!$-1$\!\!&\!\!$-1$\!\!&\!\!$-1$\!\!&\!\!$-1$\!\!&\!\!$i-1$\!\!&\!\!$i-1$\!\!&\!\!$i-1$\!\!&\!\!$C\sst i$\!\\[-.3mm]
\cline{2-10}
\!\!&\!\!$-1$\!\!&\!\!$i-1$\!\!&\!\!$-1$\!\!&\!\!$-1$\!\!&\!\!$-1$\!\!&\!\!$-1$\!\!&\!\!$i-1$\!\!&\!\!$2i-1$\!\!&\!\!$C\sst i$\!\\[-.3mm]
\cline{2-10}

\!\!&\!\!$-1$\!\!&\!\!$-1$\!\!&\!\!$-1$\!\!&\!\!$z_3$\!\!&\!\!$z_4$\!\!&\!\!$z_5$\!\!&\!\!$z_6$\!\!&\!\!$z_7$\!\!&\!\!$A\sst {z_3+z_4+z_5+z_7+3}$\!\\[-.3mm]
\hline\hline

\end{tabular}
\caption{Duals $\top^*$ of tops with $F_0^*$ one of the polygons $11,12$ of figure~\ref{fig:16pol}.}\label{tab:results1112}
\end{footnotesize}
%\end{adjustwidth}
\end{center}
\end{table}

\newpage
${}$
\vfill

\begin{table}[!b]
\begin{center}
%\beginadjustwidth}{-0.2in}{-0.2in}
\begin{footnotesize}
\begin{tabular}{|c|ccccccccc|c|}
\hline
%$F_0^*$\!\!\!&\!\!\!\multicolumn{9}{c}{Minimal points}\vline\!\!\!&\!\!\!Lie\!\\[0mm]
%\cline{2-10}
$F_0^*$\!\!\!&\!\!\!$z_0$\!\!\!&\!\!\!$z_1$\!\!\!&\!\!\!$z_2$\!\!\!&\!\!\!$z_3$\!\!\!&\!\!\!$z_4$\!\!\!&\!\!\!$z_5$\!\!\!&\!\!\!$z_6$\!\!\!&\!\!\!$z_7$\!\!\!&\!\!\!$z_8$\!\!\!&\!\!\!AKMA\!\\[0mm]
\hline\hline

13\!\!\!&\!\!\!$-1/6$\!\!\!&\!\!\!$3/2$\!\!\!&\!\!\!$-1$\!\!\!&\!\!\!$-3/2$\!\!\!&\!\!\!$-2$\!\!\!&\!\!\!$-1$\!\!\!&\!\!\!$2/3$\!\!\!&\!\!\!$7/3$\!\!\!&\!\!\!$4$\!\!\!&\!\!\!$E\sst 8$\!\\[0mm]
\cline{2-11}
\!\!\!&\!\!\!$-1/6$\!\!\!&\!\!\!$3/2$\!\!\!&\!\!\!$-1$\!\!\!&\!\!\!$-1$\!\!\!&\!\!\!$-1$\!\!\!&\!\!\!$-1$\!\!\!&\!\!\!$2/3$\!\!\!&\!\!\!$7/3$\!\!\!&\!\!\!$4$\!\!\!&\!\!\!$E\sst 8$\!\\[0mm]
\cline{2-11}
\!\!\!&\!\!\!$-1/4$\!\!\!&\!\!\!$1/2$\!\!\!&\!\!\!$2$\!\!\!&\!\!\!$-1$\!\!\!&\!\!\!$-4$\!\!\!&\!\!\!$-13/4$\!\!\!&\!\!\!$-5/2$\!\!\!&\!\!\!$-7/4$\!\!\!&\!\!\!$-1$\!\!\!&\!\!\!$E\sst 6^{(2)}$\!\\[0mm]
\cline{2-11}
\!\!\!&\!\!\!$-1/4$\!\!\!&\!\!\!$3/2$\!\!\!&\!\!\!$-1$\!\!\!&\!\!\!$-3/2$\!\!\!&\!\!\!$-2$\!\!\!&\!\!\!$-1$\!\!\!&\!\!\!$1/2$\!\!\!&\!\!\!$2$\!\!\!&\!\!\!$4$\!\!\!&\!\!\!$E\sst 7$\!\\[0mm]
\cline{2-11}
\!\!\!&\!\!\!$-1/4$\!\!\!&\!\!\!$3/2$\!\!\!&\!\!\!$-1$\!\!\!&\!\!\!$-1$\!\!\!&\!\!\!$-1$\!\!\!&\!\!\!$-1$\!\!\!&\!\!\!$1/2$\!\!\!&\!\!\!$2$\!\!\!&\!\!\!$4$\!\!\!&\!\!\!$E\sst 7$\!\\[0mm]
\cline{2-11}
\!\!\!&\!\!\!$-1/4$\!\!\!&\!\!\!$2$\!\!\!&\!\!\!$-1$\!\!\!&\!\!\!$-1$\!\!\!&\!\!\!$-1$\!\!\!&\!\!\!$-1$\!\!\!&\!\!\!$1/2$\!\!\!&\!\!\!$2$\!\!\!&\!\!\!$5$\!\!\!&\!\!\!$E\sst 7$\!\\[0mm]
\cline{2-11}
\!\!\!&\!\!\!$-1/4$\!\!\!&\!\!\!$1/2$\!\!\!&\!\!\!$2$\!\!\!&\!\!\!$-1$\!\!\!&\!\!\!$-3$\!\!\!&\!\!\!$-3$\!\!\!&\!\!\!$-7/3$\!\!\!&\!\!\!$-5/3$\!\!\!&\!\!\!$-1$\!\!\!&\!\!\!$E\sst 7$\!\\[0mm]
\cline{2-11}
\!\!\!&\!\!\!$-1/4$\!\!\!&\!\!\!$1/2$\!\!\!&\!\!\!$2$\!\!\!&\!\!\!$-1$\!\!\!&\!\!\!$-3$\!\!\!&\!\!\!$-5/2$\!\!\!&\!\!\!$-2$\!\!\!&\!\!\!$-3/2$\!\!\!&\!\!\!$-1$\!\!\!&\!\!\!$E\sst 7$\!\\[0mm]
\cline{2-11}
\!\!\!&\!\!\!$-1/4$\!\!\!&\!\!\!$1/2$\!\!\!&\!\!\!$2$\!\!\!&\!\!\!$-1$\!\!\!&\!\!\!$-2$\!\!\!&\!\!\!$-2$\!\!\!&\!\!\!$-2$\!\!\!&\!\!\!$-3/2$\!\!\!&\!\!\!$-1$\!\!\!&\!\!\!$E\sst 7$\!\\[0mm]
\cline{2-11}
\!\!\!&\!\!\!$-1/4$\!\!\!&\!\!\!$1/2$\!\!\!&\!\!\!$2$\!\!\!&\!\!\!$-1$\!\!\!&\!\!\!$-1$\!\!\!&\!\!\!$-1$\!\!\!&\!\!\!$-1$\!\!\!&\!\!\!$-1$\!\!\!&\!\!\!$-1$\!\!\!&\!\!\!$E\sst 7$\!\\[0mm]
\cline{2-11}
\!\!\!&\!\!\!$-1/3$\!\!\!&\!\!\!$1/2$\!\!\!&\!\!\!$-1$\!\!\!&\!\!\!$-1$\!\!\!&\!\!\!$-1$\!\!\!&\!\!\!$-1/3$\!\!\!&\!\!\!$1/3$\!\!\!&\!\!\!$1$\!\!\!&\!\!\!$2$\!\!\!&\!\!\!$F\sst 4$\!\\[0mm]
\cline{2-11}
\!\!\!&\!\!\!$-1/3$\!\!\!&\!\!\!$1$\!\!\!&\!\!\!$-1$\!\!\!&\!\!\!$-1$\!\!\!&\!\!\!$-1$\!\!\!&\!\!\!$-1/3$\!\!\!&\!\!\!$1/3$\!\!\!&\!\!\!$1$\!\!\!&\!\!\!$3$\!\!\!&\!\!\!$F\sst 4$\!\\[0mm]
\cline{2-11}
\!\!\!&\!\!\!$-1/3$\!\!\!&\!\!\!$1/2$\!\!\!&\!\!\!$-1$\!\!\!&\!\!\!$-1$\!\!\!&\!\!\!$0$\!\!\!&\!\!\!$0$\!\!\!&\!\!\!$1/2$\!\!\!&\!\!\!$1$\!\!\!&\!\!\!$2$\!\!\!&\!\!\!$E\sst 6$\!\\[0mm]
\cline{2-11}
\!\!\!&\!\!\!$-1/3$\!\!\!&\!\!\!$1/2$\!\!\!&\!\!\!$-1$\!\!\!&\!\!\!$-1$\!\!\!&\!\!\!$1$\!\!\!&\!\!\!$1$\!\!\!&\!\!\!$1$\!\!\!&\!\!\!$1$\!\!\!&\!\!\!$2$\!\!\!&\!\!\!$E\sst 6$\!\\[0mm]
\cline{2-11}
\!\!\!&\!\!\!$-1/3$\!\!\!&\!\!\!$1$\!\!\!&\!\!\!$-1$\!\!\!&\!\!\!$-1$\!\!\!&\!\!\!$0$\!\!\!&\!\!\!$0$\!\!\!&\!\!\!$1/2$\!\!\!&\!\!\!$1$\!\!\!&\!\!\!$3$\!\!\!&\!\!\!$E\sst 6$\!\\[0mm]
\cline{2-11}
\!\!\!&\!\!\!$-1/3$\!\!\!&\!\!\!$1$\!\!\!&\!\!\!$-1$\!\!\!&\!\!\!$-1$\!\!\!&\!\!\!$1$\!\!\!&\!\!\!$1$\!\!\!&\!\!\!$1$\!\!\!&\!\!\!$1$\!\!\!&\!\!\!$3$\!\!\!&\!\!\!$E\sst 6$\!\\[0mm]

\cline{2-11}
\!\!\!&\!\!\!$-1/2$\!\!\!&\!\!\!$i$\!\!\!&\!\!\!$-1$\!\!\!&\!\!\!$-1$\!\!\!&\!\!\!$0$\!\!\!&\!\!\!$0$\!\!\!&\!\!\!$0$\!\!\!&\!\!\!$i+1$\!\!\!&\!\!\!$2i+2$\!\!\!&\!\!\!$D\sst {2i+4}\yy D\sst 4^a\xx$\!\\[0mm]
\cline{2-11}
\!\!\!&\!\!\!$-1/2$\!\!\!&\!\!\!$i+1$\!\!\!&\!\!\!$-1$\!\!\!&\!\!\!$-1$\!\!\!&\!\!\!$-1$\!\!\!&\!\!\!$-1$\!\!\!&\!\!\!$0$\!\!\!&\!\!\!$i+1$\!\!\!&\!\!\!$2i+3$\!\!\!&\!\!\!$D\sst {2i+4}\yy B\sst 3\xx$\!\\[0mm]
\cline{2-11}
\!\!\!&\!\!\!$-1/2$\!\!\!&\!\!\!$-1$\!\!\!&\!\!\!$-1$\!\!\!&\!\!\!$0$\!\!\!&\!\!\!$i+1$\!\!\!&\!\!\!$i+1/2$\!\!\!&\!\!\!$i$\!\!\!&\!\!\!$i-1/2$\!\!\!&\!\!\!$i-1$\!\!\!&\!\!\!$D\sst {2i+4}\yy D\sst 3^{(2)a}\xx$\!\\[0mm]
\cline{2-11}
\!\!\!&\!\!\!$-1/2$\!\!\!&\!\!\!$-1$\!\!\!&\!\!\!$-1$\!\!\!&\!\!\!$0$\!\!\!&\!\!\!$i+1$\!\!\!&\!\!\!$i+1$\!\!\!&\!\!\!$i+1$\!\!\!&\!\!\!$i+1$\!\!\!&\!\!\!$i+1$\!\!\!&\!\!\!$D\sst {2i+4}\yy B\sst 3\xx$\!\\[0mm]
\cline{2-11}
\!\!\!&\!\!\!$-1/2$\!\!\!&\!\!\!$-1$\!\!\!&\!\!\!$-1$\!\!\!&\!\!\!$0$\!\!\!&\!\!\!$i+2$\!\!\!&\!\!\!$i+1$\!\!\!&\!\!\!$i$\!\!\!&\!\!\!$i$\!\!\!&\!\!\!$i$\!\!\!&\!\!\!$D\sst {2i+4}\yy D\sst 4^a\xx$\!\\[0mm]
\cline{2-11}
\!\!\!&\!\!\!$-1/2$\!\!\!&\!\!\!$-1$\!\!\!&\!\!\!$-1$\!\!\!&\!\!\!$0$\!\!\!&\!\!\!$i+1$\!\!\!&\!\!\!$i+1/2$\!\!\!&\!\!\!$i$\!\!\!&\!\!\!$i$\!\!\!&\!\!\!$i$\!\!\!&\!\!\!$D\sst {2i+4}\yy B\sst 3^a\xx$\!\\[0mm]
\cline{2-11}
\!\!\!&\!\!\!$-1/2$\!\!\!&\!\!\!$i+1$\!\!\!&\!\!\!$-1$\!\!\!&\!\!\!$-1$\!\!\!&\!\!\!$0$\!\!\!&\!\!\!$0$\!\!\!&\!\!\!$0$\!\!\!&\!\!\!$i+1$\!\!\!&\!\!\!$2i+3$\!\!\!&\!\!\!$D\sst {2i+5}$\!\\[0mm]
\cline{2-11}
\!\!\!&\!\!\!$-1/2$\!\!\!&\!\!\!$-1$\!\!\!&\!\!\!$-1$\!\!\!&\!\!\!$0$\!\!\!&\!\!\!$i+2$\!\!\!&\!\!\!$i+1$\!\!\!&\!\!\!$i+1/2$\!\!\!&\!\!\!$i$\!\!\!&\!\!\!$i$\!\!\!&\!\!\!$D\sst {2i+5}$\!\\[0mm]
\cline{2-11}
\!\!\!&\!\!\!$-1/2$\!\!\!&\!\!\!$-1$\!\!\!&\!\!\!$-1$\!\!\!&\!\!\!$0$\!\!\!&\!\!\!$i+2$\!\!\!&\!\!\!$i+1$\!\!\!&\!\!\!$i+1$\!\!\!&\!\!\!$i+1$\!\!\!&\!\!\!$i+1$\!\!\!&\!\!\!$D\sst {2i+5}$\!\\[0mm]
\cline{2-11}

\!\!\!&\!\!\!$-1/2$\!\!\!&\!\!\!$i/2$\!\!\!&\!\!\!$-1$\!\!\!&\!\!\!$-1$\!\!\!&\!\!\!$0$\!\!\!&\!\!\!$0$\!\!\!&\!\!\!$0$\!\!\!&\!\!\!$(i+1)/2$\!\!\!&\!\!\!$i+1$\!\!\!&\!\!\!$B\sst {i+3}\yy B\sst 3^a\xx$\!\\[0mm]
\cline{2-11}
\!\!\!&\!\!\!$-1/2$\!\!\!&\!\!\!$(i+1)/2$\!\!\!&\!\!\!$-1$\!\!\!&\!\!\!$-1$\!\!\!&\!\!\!$-1$\!\!\!&\!\!\!$-1$\!\!\!&\!\!\!$0$\!\!\!&\!\!\!$i/2+1$\!\!\!&\!\!\!$i+2$\!\!\!&\!\!\!$B\sst {i+3}\yy G\sst 2\xx$\!\\[0mm]
\cline{2-11}
\!\!\!&\!\!\!$-1/2$\!\!\!&\!\!\!$i+1/2$\!\!\!&\!\!\!$-1$\!\!\!&\!\!\!$-1$\!\!\!&\!\!\!$-1$\!\!\!&\!\!\!$-1/2$\!\!\!&\!\!\!$0$\!\!\!&\!\!\!$i+1$\!\!\!&\!\!\!$2i+2$\!\!\!&\!\!\!$D\sst {2i+4}^{(2)}$\!\\[0mm]
\cline{2-11}
\!\!\!&\!\!\!$-1/2$\!\!\!&\!\!\!$i$\!\!\!&\!\!\!$-1$\!\!\!&\!\!\!$-1$\!\!\!&\!\!\!$-1$\!\!\!&\!\!\!$-1/2$\!\!\!&\!\!\!$0$\!\!\!&\!\!\!$i+1/2$\!\!\!&\!\!\!$2i+1$\!\!\!&\!\!\!$D\sst {2i+3}^{(2)}\yy D\sst 3^{(2)a}\xx$\!\\[0mm]
\cline{2-11}
\!\!\!&\!\!\!$-1/2$\!\!\!&\!\!\!$i-1/2$\!\!\!&\!\!\!$-1$\!\!\!&\!\!\!$-1/2$\!\!\!&\!\!\!$0$\!\!\!&\!\!\!$0$\!\!\!&\!\!\!$0$\!\!\!&\!\!\!$i$\!\!\!&\!\!\!$2i$\!\!\!&\!\!\!$D\sst
{2i+3}^{(2)}\yy A\sst 2^{(2)}\xx$\!\\[0mm]
\cline{2-11}

\!\!\!&\!\!\!$-1$\!\!\!&\!\!\!$i-1$\!\!\!&\!\!\!$-1$\!\!\!&\!\!\!$-1$\!\!\!&\!\!\!$-1$\!\!\!&\!\!\!$-1$\!\!\!&\!\!\!$-1$\!\!\!&\!\!\!$i-1$\!\!\!&\!\!\!$2i-1$\!\!\!&\!\!\!$C\sst i$\!\\[0mm]
\cline{2-11}
\!\!\!&\!\!\!$-1$\!\!\!&\!\!\!$-1$\!\!\!&\!\!\!$-1$\!\!\!&\!\!\!$-1$\!\!\!&\!\!\!$i-1$\!\!\!&\!\!\!$i-1$\!\!\!&\!\!\!$i-1$\!\!\!&\!\!\!$i-1$\!\!\!&\!\!\!$i-1$\!\!\!&\!\!\!$C\sst i$\!\\[0mm]
\cline{2-11}

\!\!\!&\!\!\!$-1$\!\!\!&\!\!\!$-1$\!\!\!&\!\!\!$-1$\!\!\!&\!\!\!$z_3$\!\!\!&\!\!\!$z_4$\!\!\!&\!\!\!$z_5$\!\!\!&\!\!\!$z_6$\!\!\!&\!\!\!$z_7$\!\!\!&\!\!\!$z_8$\!\!\!&\!\!\!$A\sst {z_4+z_8+1}$\!\\[0mm]

\hline\hline

\end{tabular}
\caption{Duals $\top^*$ of tops with $F_0^*$ being polygon $13$ of figure~\ref{fig:16pol}.}\label{tab:results13}
\end{footnotesize}
\end{center}
\end{table}

\newpage

${}$
\vfill

\begin{table}[!b]
\begin{center}
\begin{footnotesize}
\begin{tabular}{|c|ccccccccc|c|}
\hline
%$F_0^*$\!&\!\multicolumn{9}{c}{Minimal points}\vline\!&\!Lie\!\\[0mm]
%\cline{2-10}\!\!
$\!F_0^*$\!&\!$z_0$\!&\!$z_1$\!&\!$z_2$\!&\!$z_3$\!&\!$z_4$\!&\!$z_5$\!&\!$z_6$\!&\!$z_7$\!&\!$z_8$\!&\!AKMA\!\\[0mm]
\hline\hline

\!14\!&\!$-1/6$\!&\!$-1$\!&\!$-2$\!&\!$-3/2$\!&\!$-1$\!&\!$2/3$\!&\!$7/3$\!&\!$4$\!&\!$3/2$\!&\!$E\sst 8$\!\\[0mm]
\cline{2-11}
\!&\!$-1/6$\!&\!$-1$\!&\!$-1$\!&\!$-1$\!&\!$-1$\!&\!$2/3$\!&\!$7/3$\!&\!$4$\!&\!$3/2$\!&\!$E\sst 8$\!\\[0mm]
\cline{2-11}
\!&\!$-1/4$\!&\!$2$\!&\!$2$\!&\!$1/2$\!&\!$-1$\!&\!$-5/3$\!&\!$-7/3$\!&\!$-3$\!&\!$-1$\!&\!$E\sst 7$\!\\[0mm]
\cline{2-11}
\!&\!$-1/4$\!&\!$2$\!&\!$2$\!&\!$1/2$\!&\!$-1$\!&\!$-3/2$\!&\!$-2$\!&\!$-2$\!&\!$-1$\!&\!$E\sst 7$\!\\[0mm]
\cline{2-11}
\!&\!$-1/4$\!&\!$2$\!&\!$2$\!&\!$1/2$\!&\!$-1$\!&\!$-1$\!&\!$-1$\!&\!$-1$\!&\!$-1$\!&\!$E\sst 7$\!\\[0mm]
\cline{2-11}
\!&\!$-1/4$\!&\!$-1$\!&\!$-2$\!&\!$-3/2$\!&\!$-1$\!&\!$1/2$\!&\!$2$\!&\!$4$\!&\!$3/2$\!&\!$E\sst 7$\!\\[0mm]
\cline{2-11}
\!&\!$-1/4$\!&\!$-1$\!&\!$-2$\!&\!$-3/2$\!&\!$-1$\!&\!$1/2$\!&\!$2$\!&\!$5$\!&\!$2$\!&\!$E\sst 7$\!\\[0mm]
\cline{2-11}
\!&\!$-1/4$\!&\!$-1$\!&\!$-1$\!&\!$-1$\!&\!$-1$\!&\!$1/2$\!&\!$2$\!&\!$4$\!&\!$3/2$\!&\!$E\sst 7$\!\\[0mm]
\cline{2-11}
\!&\!$-1/4$\!&\!$-1$\!&\!$-1$\!&\!$-1$\!&\!$-1$\!&\!$1/2$\!&\!$2$\!&\!$5$\!&\!$2$\!&\!$E\sst 7$\!\\[0mm]
\cline{2-11}
\!&\!$-1/3$\!&\!$-1$\!&\!$0$\!&\!$1/2$\!&\!$1$\!&\!$1/3$\!&\!$-1/3$\!&\!$-1$\!&\!$-1$\!&\!$F\sst 4$\!\\[0mm]
\cline{2-11}
\!&\!$-1/3$\!&\!$-1$\!&\!$1$\!&\!$1$\!&\!$1$\!&\!$1/3$\!&\!$-1/3$\!&\!$-1$\!&\!$-1$\!&\!$F\sst 4$\!\\[0mm]
\cline{2-11}
\!&\!$-1/3$\!&\!$-1$\!&\!$0$\!&\!$1/2$\!&\!$1$\!&\!$1/2$\!&\!$0$\!&\!$0$\!&\!$-1$\!&\!$E\sst 6$\!\\[0mm]
\cline{2-11}
\!&\!$-1/3$\!&\!$-1$\!&\!$0$\!&\!$1/2$\!&\!$1$\!&\!$1$\!&\!$1$\!&\!$1$\!&\!$-1$\!&\!$E\sst 6$\!\\[0mm]
\cline{2-11}
\!&\!$-1/3$\!&\!$-1$\!&\!$1$\!&\!$1$\!&\!$1$\!&\!$1/2$\!&\!$0$\!&\!$0$\!&\!$-1$\!&\!$E\sst 6$\!\\[0mm]
\cline{2-11}
\!&\!$-1/3$\!&\!$-1$\!&\!$1$\!&\!$1$\!&\!$1$\!&\!$1$\!&\!$1$\!&\!$1$\!&\!$-1$\!&\!$E\sst 6$\!\\[0mm]
\cline{2-11}
\!&\!$-1/3$\!&\!$-1$\!&\!$-1$\!&\!$-1$\!&\!$0$\!&\!$1/2$\!&\!$1$\!&\!$2$\!&\!$1/2$\!&\!$E\sst 6$\!\\[0mm]
\cline{2-11}
\!&\!$-1/3$\!&\!$-1$\!&\!$-1$\!&\!$-1$\!&\!$1$\!&\!$1$\!&\!$1$\!&\!$2$\!&\!$1/2$\!&\!$E\sst 6$\!\\[0mm]
\cline{2-11}
\!&\!$-1/3$\!&\!$-1$\!&\!$-1$\!&\!$-1$\!&\!$0$\!&\!$1/2$\!&\!$1$\!&\!$3$\!&\!$1$\!&\!$E\sst 6$\!\\[0mm]
\cline{2-11}
\!&\!$-1/3$\!&\!$-1$\!&\!$-1$\!&\!$-1$\!&\!$1$\!&\!$1$\!&\!$1$\!&\!$3$\!&\!$1$\!&\!$E\sst 6$\!\\[0mm]
\cline{2-11}
\!&\!$-1/2$\!&\!$-1$\!&\!$-1$\!&\!$-1$\!&\!$i$\!&\!$i$\!&\!$i+1/2$\!&\!$i+1$\!&\!$0$\!&\!$D\sst {2i+4}\yy B\sst 3^a\xx$\!\\[0mm]
\cline{2-11}
\!&\!$-1/2$\!&\!$-1$\!&\!$-1$\!&\!$-1$\!&\!$i$\!&\!$i$\!&\!$i+1$\!&\!$i+2$\!&\!$0$\!&\!$D\sst {2i+4}\yy D\sst 4^a\xx$\!\\[0mm]
\cline{2-11}
\!&\!$-1/2$\!&\!$-1$\!&\!$-1$\!&\!$-1$\!&\!$i+1$\!&\!$i+1$\!&\!$i+1$\!&\!$i+1$\!&\!$0$\!&\!$D\sst {2i+4}\yy B\sst 3\xx$\!\\[0mm]
\cline{2-11}
\!&\!$-1/2$\!&\!$-1$\!&\!$-1$\!&\!$-1/2$\!&\!$0$\!&\!$0$\!&\!$i$\!&\!$2i+1$\!&\!$i$\!&\!$D\sst {2i+4}\yy B\sst 3^b\xx$\!\\[0mm]
\cline{2-11}
\!&\!$-1/2$\!&\!$-1$\!&\!$-1$\!&\!$0$\!&\!$1$\!&\!$0$\!&\!$i$\!&\!$2i+1$\!&\!$i$\!&\!$D\sst {2i+4}$\!\\[0mm]
\cline{2-11}
\!&\!$-1/2$\!&\!$-1$\!&\!$-1$\!&\!$-1$\!&\!$0$\!&\!$0$\!&\!$i+1$\!&\!$2i+2$\!&\!$i$\!&\!$D\sst {2i+4}\yy D\sst 4^a\xx$\!\\[0mm]
\cline{2-11}
\!&\!$-1/2$\!&\!$-1$\!&\!$-1$\!&\!$-1$\!&\!$-1$\!&\!$0$\!&\!$i+1$\!&\!$2i+3$\!&\!$i+1$\!&\!$D\sst {2i+4}\yy B\sst 3\xx$\!\\[0mm]
\cline{2-11}
\!&\!$-1/2$\!&\!$-1$\!&\!$-1$\!&\!$-1$\!&\!$i$\!&\!$i+1/2$\!&\!$i+1$\!&\!$i+2$\!&\!$0$\!&\!$D\sst {2i+5}$\!\\[0mm]
\cline{2-11}
\!&\!$-1/2$\!&\!$-1$\!&\!$-1$\!&\!$-1$\!&\!$i$\!&\!$i+1$\!&\!$i+2$\!&\!$i+3$\!&\!$0$\!&\!$D\sst {2i+5}$\!\\[0mm]
\cline{2-11}
\!&\!$-1/2$\!&\!$-1$\!&\!$-1$\!&\!$-1$\!&\!$i+1$\!&\!$i+1$\!&\!$i+1$\!&\!$i+2$\!&\!$0$\!&\!$D\sst {2i+5}$\!\\[0mm]
\cline{2-11}
\!&\!$-1/2$\!&\!$-1$\!&\!$-1$\!&\!$-1/2$\!&\!$0$\!&\!$0$\!&\!$i+1$\!&\!$2i+2$\!&\!$i$\!&\!$D\sst {2i+5}$\!\\[0mm]
\cline{2-11}
\!&\!$-1/2$\!&\!$-1$\!&\!$-1$\!&\!$0$\!&\!$1$\!&\!$0$\!&\!$i+1$\!&\!$2i+2$\!&\!$i$\!&\!$D\sst {2i+5}$\!\\[0mm]
\cline{2-11}
\!&\!$-1/2$\!&\!$-1$\!&\!$-1$\!&\!$-1$\!&\!$0$\!&\!$0$\!&\!$i+1$\!&\!$2i+3$\!&\!$i+1$\!&\!$D\sst {2i+5}$\!\\[0mm]
\cline{2-11}
\!&\!$-1/2$\!&\!$-1$\!&\!$-1$\!&\!$-1$\!&\!$-1$\!&\!$0$\!&\!$i+2$\!&\!$2i+4$\!&\!$i+1$\!&\!$D\sst {2i+5}$\!\\[0mm]
\cline{2-11}
\!&\!$-1/2$\!&\!$-1$\!&\!$-1$\!&\!$-1/2$\!&\!$0$\!&\!$0$\!&\!$i/2$\!&\!$i$\!&\!$(i-1)/2$\!&\!$B\sst {i+3}\yy G\sst 2\xx$\!\\[0mm]
\cline{2-11}
\!&\!$-1/2$\!&\!$-1$\!&\!$-1$\!&\!$0$\!&\!$1$\!&\!$0$\!&\!$i/2$\!&\!$i$\!&\!$(i-1)/2$\!&\!$B\sst {i+3}\yy B\sst 3^b\xx$\!\\[0mm]
\cline{2-11}
\!&\!$-1/2$\!&\!$-1$\!&\!$-1$\!&\!$-1$\!&\!$0$\!&\!$0$\!&\!$(i+1)/2$\!&\!$i+1$\!&\!$i/2$\!&\!$B\sst {i+3}\yy B\sst 3^a\xx$\!\\[0mm]
\cline{2-11}
\!&\!$-1/2$\!&\!$-1$\!&\!$-1$\!&\!$-1$\!&\!$-1$\!&\!$0$\!&\!$i/2+1$\!&\!$i+2$\!&\!$(i+1)/2$\!&\!$B\sst {i+3}\yy G\sst 2\xx$\!\\[0mm]
\cline{2-11}

\!&\!$-1$\!&\!$-1$\!&\!$-1$\!&\!$-1$\!&\!$-1$\!&\!$-1$\!&\!$i-1$\!&\!$2i-1$\!&\!$i-1$\!&\!$C\sst i$\!\\[0mm]
\cline{2-11}
\!&\!$-1$\!&\!$-1$\!&\!$-1$\!&\!$-1$\!&\!$i-1$\!&\!$i-1$\!&\!$i-1$\!&\!$i-1$\!&\!$-1$\!&\!$C\sst i$\!\\[0mm]
\cline{2-11}

\!&\!$-1$\!&\!$-1$\!&\!$-1$\!&\!$z_3$\!&\!$z_4$\!&\!$z_5$\!&\!$z_6$\!&\!$z_7$\!&\!$z_8$\!&\!$A\sst {z_4+z_7+1}$\!\\[0mm]

\hline\hline

\end{tabular}
\caption{Duals $\top^*$ of tops with $F_0^*$ being polygon $14$ of figure~\ref{fig:16pol}.}\label{tab:results14}
\end{footnotesize}
\end{center}
\end{table}

\newpage
%${}$
\vspace*{\fill}

\begin{table}[!b]
\begin{center}
\begin{footnotesize}
\begin{tabular}{|c|cccccccccc|c|}
\hline
%$F_0^*$\!\!\!&\!\!\!\multicolumn{10}{c}{Minimal points}\vline\!\!\!&\!\!\!Lie\!\\[-.4mm]
%\cline{2-11}
$F_0^*$\!\!\!&\!\!\!$z_0$\!\!\!&\!\!\!$z_1$\!\!\!&\!\!\!$z_2$\!\!\!&\!\!\!$z_3$\!\!\!&\!\!\!$z_4$\!\!\!&\!\!\!$z_5$\!\!\!&\!\!\!$z_6$\!\!\!&\!\!\!$z_7$\!\!\!&\!\!\!$z_8$\!\!\!&\!\!\!$z_9$\!\!\!&\!\!\!AKMA\!\\[-.4mm]
\hline\hline

15\!\!\!&\!\!\!$-1/4$\!\!\!&\!\!\!$-1$\!\!\!&\!\!\!$1$\!\!\!&\!\!\!$3/2$\!\!\!&\!\!\!$2$\!\!\!&\!\!\!$1/2$\!\!\!&\!\!\!$-1$\!\!\!&\!\!\!$-3/2$\!\!\!&\!\!\!$-2$\!\!\!&\!\!\!-\!\!\!&\!\!\!$E\sst 7$\!\\[-.4mm]
\cline{2-12}
\!\!\!&\!\!\!$-1/4$\!\!\!&\!\!\!$-1$\!\!\!&\!\!\!$1$\!\!\!&\!\!\!$3/2$\!\!\!&\!\!\!$2$\!\!\!&\!\!\!$1/2$\!\!\!&\!\!\!$-1$\!\!\!&\!\!\!$-1$\!\!\!&\!\!\!$-1$\!\!\!&\!\!\!-\!\!\!&\!\!\!$E\sst 7$\!\\[-.4mm]
\cline{2-12}
\!\!\!&\!\!\!$-1/4$\!\!\!&\!\!\!$-1$\!\!\!&\!\!\!$2$\!\!\!&\!\!\!$2$\!\!\!&\!\!\!$2$\!\!\!&\!\!\!$1/2$\!\!\!&\!\!\!$-1$\!\!\!&\!\!\!$-1$\!\!\!&\!\!\!$-1$\!\!\!&\!\!\!-\!\!\!&\!\!\!$E\sst 7$\!\\[-.4mm]
\cline{2-12}
\!\!\!&\!\!\!$-1/3$\!\!\!&\!\!\!$-1$\!\!\!&\!\!\!$-1$\!\!\!&\!\!\!$-1$\!\!\!&\!\!\!$0$\!\!\!&\!\!\!$1/2$\!\!\!&\!\!\!$1$\!\!\!&\!\!\!$1/2$\!\!\!&\!\!\!$0$\!\!\!&\!\!\!-\!\!\!&\!\!\!$E\sst 6$\!\\[-.4mm]
\cline{2-12}
\!\!\!&\!\!\!$-1/3$\!\!\!&\!\!\!$-1$\!\!\!&\!\!\!$-1$\!\!\!&\!\!\!$-1$\!\!\!&\!\!\!$0$\!\!\!&\!\!\!$1/2$\!\!\!&\!\!\!$1$\!\!\!&\!\!\!$1$\!\!\!&\!\!\!$1$\!\!\!&\!\!\!-\!\!\!&\!\!\!$E\sst 6$\!\\[-.4mm]
\cline{2-12}
\!\!\!&\!\!\!$-1/3$\!\!\!&\!\!\!$-1$\!\!\!&\!\!\!$-1$\!\!\!&\!\!\!$-1$\!\!\!&\!\!\!$1$\!\!\!&\!\!\!$1$\!\!\!&\!\!\!$1$\!\!\!&\!\!\!$1$\!\!\!&\!\!\!$1$\!\!\!&\!\!\!-\!\!\!&\!\!\!$E\sst 6$\!\\[-.4mm]
\cline{2-12}

\!\!\!&\!\!\!$-1/2$\!\!\!&\!\!\!$-1$\!\!\!&\!\!\!$-1$\!\!\!&\!\!\!$-1/2$\!\!\!&\!\!\!$0$\!\!\!&\!\!\!$0$\!\!\!&\!\!\!$i$\!\!\!&\!\!\!$i-1/2$\!\!\!&\!\!\!$i-1$\!\!\!&\!\!\!-\!\!\!&\!\!\!$D\sst {2i+4}\yy D\sst 3^{(2)a}\xx$\!\\[-.4mm]
\cline{2-12}
\!\!\!&\!\!\!$-1/2$\!\!\!&\!\!\!$-1$\!\!\!&\!\!\!$-1$\!\!\!&\!\!\!$-1/2$\!\!\!&\!\!\!$0$\!\!\!&\!\!\!$0$\!\!\!&\!\!\!$i$\!\!\!&\!\!\!$i$\!\!\!&\!\!\!$i$\!\!\!&\!\!\!-\!\!\!&\!\!\!$D\sst {2i+4}\yy B\sst 3^a\xx$\!\\[-.4mm]
\cline{2-12}
\!\!\!&\!\!\!$-1/2$\!\!\!&\!\!\!$-1$\!\!\!&\!\!\!$-1$\!\!\!&\!\!\!$0$\!\!\!&\!\!\!$1$\!\!\!&\!\!\!$0$\!\!\!&\!\!\!$i$\!\!\!&\!\!\!$i$\!\!\!&\!\!\!$i$\!\!\!&\!\!\!-\!\!\!&\!\!\!$D\sst {2i+4}\yy D\sst 4^a\xx$\!\\[-.4mm]
\cline{2-12}
\!\!\!&\!\!\!$-1/2$\!\!\!&\!\!\!$-1$\!\!\!&\!\!\!$-1$\!\!\!&\!\!\!$0$\!\!\!&\!\!\!$1$\!\!\!&\!\!\!$0$\!\!\!&\!\!\!$i+1$\!\!\!&\!\!\!$i$\!\!\!&\!\!\!$i-1$\!\!\!&\!\!\!-\!\!\!&\!\!\!$D\sst {2i+4}\yy B\sst 3\xx$\!\\[-.4mm]
\cline{2-12}
\!\!\!&\!\!\!$-1/2$\!\!\!&\!\!\!$-1$\!\!\!&\!\!\!$-1$\!\!\!&\!\!\!$-1$\!\!\!&\!\!\!$0$\!\!\!&\!\!\!$0$\!\!\!&\!\!\!$i+1$\!\!\!&\!\!\!$i$\!\!\!&\!\!\!$i$\!\!\!&\!\!\!-\!\!\!&\!\!\!$D\sst {2i+4}$\!\\[-.4mm]
\cline{2-12}
\!\!\!&\!\!\!$-1/2$\!\!\!&\!\!\!$i$\!\!\!&\!\!\!$-1$\!\!\!&\!\!\!$-1$\!\!\!&\!\!\!$0$\!\!\!&\!\!\!$0$\!\!\!&\!\!\!$0$\!\!\!&\!\!\!$i$\!\!\!&\!\!\!$2i+1$\!\!\!&\!\!\!-\!\!\!&\!\!\!$D\sst {2i+4}\yy D\sst 4^a\xx$\!\\[-.4mm]
\cline{2-12}
\!\!\!&\!\!\!$-1/2$\!\!\!&\!\!\!$-1$\!\!\!&\!\!\!$-1$\!\!\!&\!\!\!$-1/2$\!\!\!&\!\!\!$0$\!\!\!&\!\!\!$0$\!\!\!&\!\!\!$i+1$\!\!\!&\!\!\!$i$\!\!\!&\!\!\!$i$\!\!\!&\!\!\!-\!\!\!&\!\!\!$D\sst {2i+5}$\!\\[-.4mm]
\cline{2-12}
\!\!\!&\!\!\!$-1/2$\!\!\!&\!\!\!$-1$\!\!\!&\!\!\!$-1$\!\!\!&\!\!\!$0$\!\!\!&\!\!\!$1$\!\!\!&\!\!\!$0$\!\!\!&\!\!\!$i+1$\!\!\!&\!\!\!$i$\!\!\!&\!\!\!$i$\!\!\!&\!\!\!-\!\!\!&\!\!\!$D\sst {2i+5}$\!\\[-.4mm]
\cline{2-12}
\!\!\!&\!\!\!$-1/2$\!\!\!&\!\!\!$i$\!\!\!&\!\!\!$-1$\!\!\!&\!\!\!$-1$\!\!\!&\!\!\!$0$\!\!\!&\!\!\!$0$\!\!\!&\!\!\!$0$\!\!\!&\!\!\!$i+1$\!\!\!&\!\!\!$2i+2$\!\!\!&\!\!\!-\!\!\!&\!\!\!$D\sst {2i+5}$\!\\[-.4mm]
\cline{2-12}
\!\!\!&\!\!\!$-1/2$\!\!\!&\!\!\!$(i-1)/2$\!\!\!&\!\!\!$-1$\!\!\!&\!\!\!$-1$\!\!\!&\!\!\!$0$\!\!\!&\!\!\!$0$\!\!\!&\!\!\!$0$\!\!\!&\!\!\!$i/2$\!\!\!&\!\!\!$i$\!\!\!&\!\!\!-\!\!\!&\!\!\!$B\sst {i+3}\yy B\sst 3^a\xx$\!\\[-.4mm]
\cline{2-12}
\!\!\!&\!\!\!$-1/2$\!\!\!&\!\!\!$(i-1)/2$\!\!\!&\!\!\!$-1$\!\!\!&\!\!\!$-1$\!\!\!&\!\!\!$-1$\!\!\!&\!\!\!$-1/2$\!\!\!&\!\!\!$0$\!\!\!&\!\!\!$i/2$\!\!\!&\!\!\!$i$\!\!\!&\!\!\!-\!\!\!&\!\!\!$D\sst {i+3}^{(2)}\yy D\sst 3^{(2)a}\xx$\!\\[-.4mm]
\cline{2-12}

\!\!\!&\!\!\!$-1$\!\!\!&\!\!\!$-1$\!\!\!&\!\!\!$-1$\!\!\!&\!\!\!$-1$\!\!\!&\!\!\!$-1$\!\!\!&\!\!\!$-1$\!\!\!&\!\!\!$i-1$\!\!\!&\!\!\!$i-1$\!\!\!&\!\!\!$i-1$\!\!\!&\!\!\!-\!\!\!&\!\!\!$C\sst i$\!\\[-.4mm]
\cline{2-12}
\!\!\!&\!\!\!$-1$\!\!\!&\!\!\!$i-1$\!\!\!&\!\!\!$-1$\!\!\!&\!\!\!$-1$\!\!\!&\!\!\!$-1$\!\!\!&\!\!\!$-1$\!\!\!&\!\!\!$-1$\!\!\!&\!\!\!$i-1$\!\!\!&\!\!\!$2i-1$\!\!\!&\!\!\!-\!\!\!&\!\!\!$C\sst i$\!\\[-.4mm]
\cline{2-12}

\!\!\!&\!\!\!$-1$\!\!\!&\!\!\!$-1$\!\!\!&\!\!\!$-1$\!\!\!&\!\!\!$z_3$\!\!\!&\!\!\!$z_4$\!\!\!&\!\!\!$z_5$\!\!\!&\!\!\!$z_6$\!\!\!&\!\!\!$z_7$\!\!\!&\!\!\!$z_8$\!\!\!&\!\!\!-\!\!\!&\!\!\!$A\sst {z_4+z_6+z_8+2}$\!\\[-.4mm]
\hline\hline
16\!\!\!&\!\!\!$-1/6$\!\!\!&\!\!\!$-1$\!\!\!&\!\!\!$-3$\!\!\!&\!\!\!$-2$\!\!\!&\!\!\!$-3/2$\!\!\!&\!\!\!$-1$\!\!\!&\!\!\!$2/3$\!\!\!&\!\!\!$7/3$\!\!\!&\!\!\!$4$\!\!\!&\!\!\!$3/2$\!\!\!&\!\!\!$E\sst 8$\!\\[-.4mm]
\cline{2-12}
\!\!\!&\!\!\!$-1/6$\!\!\!&\!\!\!$-1$\!\!\!&\!\!\!$-3$\!\!\!&\!\!\!$-7/3$\!\!\!&\!\!\!$-5/3$\!\!\!&\!\!\!$-1$\!\!\!&\!\!\!$2/3$\!\!\!&\!\!\!$7/3$\!\!\!&\!\!\!$4$\!\!\!&\!\!\!$3/2$\!\!\!&\!\!\!$E\sst 8$\!\\[-.4mm]
\cline{2-12}
\!\!\!&\!\!\!$-1/6$\!\!\!&\!\!\!$-1$\!\!\!&\!\!\!$-1$\!\!\!&\!\!\!$-1$\!\!\!&\!\!\!$-1$\!\!\!&\!\!\!$-1$\!\!\!&\!\!\!$2/3$\!\!\!&\!\!\!$7/3$\!\!\!&\!\!\!$4$\!\!\!&\!\!\!$3/2$\!\!\!&\!\!\!$E\sst 8$\!\\[-.4mm]
\cline{2-12}

\!\!\!&\!\!\!$-1/4$\!\!\!&\!\!\!$-1$\!\!\!&\!\!\!$-3$\!\!\!&\!\!\!$-7/3$\!\!\!&\!\!\!$-5/3$\!\!\!&\!\!\!$-1$\!\!\!&\!\!\!$1/2$\!\!\!&\!\!\!$2$\!\!\!&\!\!\!$4$\!\!\!&\!\!\!$3/2$\!\!\!&\!\!\!$E\sst 7$\!\\[-.4mm]
\cline{2-12}
\!\!\!&\!\!\!$-1/4$\!\!\!&\!\!\!$-1$\!\!\!&\!\!\!$-3$\!\!\!&\!\!\!$-7/3$\!\!\!&\!\!\!$-5/3$\!\!\!&\!\!\!$-1$\!\!\!&\!\!\!$1/2$\!\!\!&\!\!\!$2$\!\!\!&\!\!\!$5$\!\!\!&\!\!\!$2$\!\!\!&\!\!\!$E\sst 7$\!\\[-.4mm]
\cline{2-12}
\!\!\!&\!\!\!$-1/4$\!\!\!&\!\!\!$-1$\!\!\!&\!\!\!$-2$\!\!\!&\!\!\!$-2$\!\!\!&\!\!\!$-3/2$\!\!\!&\!\!\!$-1$\!\!\!&\!\!\!$1/2$\!\!\!&\!\!\!$2$\!\!\!&\!\!\!$4$\!\!\!&\!\!\!$3/2$\!\!\!&\!\!\!$E\sst 7$\!\\[-.4mm]
\cline{2-12}
\!\!\!&\!\!\!$-1/4$\!\!\!&\!\!\!$-1$\!\!\!&\!\!\!$-2$\!\!\!&\!\!\!$-2$\!\!\!&\!\!\!$-3/2$\!\!\!&\!\!\!$-1$\!\!\!&\!\!\!$1/2$\!\!\!&\!\!\!$2$\!\!\!&\!\!\!$5$\!\!\!&\!\!\!$2$\!\!\!&\!\!\!$E\sst 7$\!\\[-.4mm]
\cline{2-12}
\!\!\!&\!\!\!$-1/4$\!\!\!&\!\!\!$-1$\!\!\!&\!\!\!$-1$\!\!\!&\!\!\!$-1$\!\!\!&\!\!\!$-1$\!\!\!&\!\!\!$-1$\!\!\!&\!\!\!$1/2$\!\!\!&\!\!\!$2$\!\!\!&\!\!\!$4$\!\!\!&\!\!\!$3/2$\!\!\!&\!\!\!$E\sst 7$\!\\[-.4mm]
\cline{2-12}
\!\!\!&\!\!\!$-1/4$\!\!\!&\!\!\!$-1$\!\!\!&\!\!\!$-1$\!\!\!&\!\!\!$-1$\!\!\!&\!\!\!$-1$\!\!\!&\!\!\!$-1$\!\!\!&\!\!\!$1/2$\!\!\!&\!\!\!$2$\!\!\!&\!\!\!$5$\!\!\!&\!\!\!$2$\!\!\!&\!\!\!$E\sst 7$\!\\[-.4mm]
\cline{2-12}
\!\!\!&\!\!\!$-1/3$\!\!\!&\!\!\!$-1$\!\!\!&\!\!\!$-1$\!\!\!&\!\!\!$-1/3$\!\!\!&\!\!\!$1/3$\!\!\!&\!\!\!$1$\!\!\!&\!\!\!$1/3$\!\!\!&\!\!\!$-1/3$\!\!\!&\!\!\!$-1$\!\!\!&\!\!\!$-1$\!\!\!&\!\!\!$D\sst 4^{(3)}$\!\\[-.4mm]
\cline{2-12}
\!\!\!&\!\!\!$-1/3$\!\!\!&\!\!\!$-1$\!\!\!&\!\!\!$-1$\!\!\!&\!\!\!$-1/3$\!\!\!&\!\!\!$1/3$\!\!\!&\!\!\!$1$\!\!\!&\!\!\!$1/2$\!\!\!&\!\!\!$0$\!\!\!&\!\!\!$0$\!\!\!&\!\!\!$-1$\!\!\!&\!\!\!$F\sst 4$\!\\[-.4mm]
\cline{2-12}
\!\!\!&\!\!\!$-1/3$\!\!\!&\!\!\!$-1$\!\!\!&\!\!\!$-1$\!\!\!&\!\!\!$-1/3$\!\!\!&\!\!\!$1/3$\!\!\!&\!\!\!$1$\!\!\!&\!\!\!$1$\!\!\!&\!\!\!$1$\!\!\!&\!\!\!$1$\!\!\!&\!\!\!$-1$\!\!\!&\!\!\!$F\sst 4$\!\\[-.4mm]
\cline{2-12}
\!\!\!&\!\!\!$-1/3$\!\!\!&\!\!\!$-1$\!\!\!&\!\!\!$0$\!\!\!&\!\!\!$0$\!\!\!&\!\!\!$1/2$\!\!\!&\!\!\!$1$\!\!\!&\!\!\!$1/2$\!\!\!&\!\!\!$0$\!\!\!&\!\!\!$0$\!\!\!&\!\!\!$-1$\!\!\!&\!\!\!$E\sst 6$\!\\[-.4mm]
\cline{2-12}
\!\!\!&\!\!\!$-1/3$\!\!\!&\!\!\!$-1$\!\!\!&\!\!\!$0$\!\!\!&\!\!\!$0$\!\!\!&\!\!\!$1/2$\!\!\!&\!\!\!$1$\!\!\!&\!\!\!$1$\!\!\!&\!\!\!$1$\!\!\!&\!\!\!$1$\!\!\!&\!\!\!$-1$\!\!\!&\!\!\!$E\sst 6$\!\\[-.4mm]
\cline{2-12}
\!\!\!&\!\!\!$-1/3$\!\!\!&\!\!\!$-1$\!\!\!&\!\!\!$1$\!\!\!&\!\!\!$1$\!\!\!&\!\!\!$1$\!\!\!&\!\!\!$1$\!\!\!&\!\!\!$1$\!\!\!&\!\!\!$1$\!\!\!&\!\!\!$1$\!\!\!&\!\!\!$-1$\!\!\!&\!\!\!$E\sst 6$\!\\[-.4mm]
\cline{2-12}
\!\!\!&\!\!\!$-1/3$\!\!\!&\!\!\!$-1$\!\!\!&\!\!\!$-2$\!\!\!&\!\!\!$-3/2$\!\!\!&\!\!\!$-1$\!\!\!&\!\!\!$0$\!\!\!&\!\!\!$1/2$\!\!\!&\!\!\!$1$\!\!\!&\!\!\!$2$\!\!\!&\!\!\!$1/2$\!\!\!&\!\!\!$E\sst 6$\!\\[-.4mm]
\cline{2-12}
\!\!\!&\!\!\!$-1/3$\!\!\!&\!\!\!$-1$\!\!\!&\!\!\!$-2$\!\!\!&\!\!\!$-3/2$\!\!\!&\!\!\!$-1$\!\!\!&\!\!\!$1$\!\!\!&\!\!\!$1$\!\!\!&\!\!\!$1$\!\!\!&\!\!\!$2$\!\!\!&\!\!\!$1/2$\!\!\!&\!\!\!$E\sst 6$\!\\[-.4mm]
\cline{2-12}
\!\!\!&\!\!\!$-1/3$\!\!\!&\!\!\!$-1$\!\!\!&\!\!\!$-2$\!\!\!&\!\!\!$-3/2$\!\!\!&\!\!\!$-1$\!\!\!&\!\!\!$1$\!\!\!&\!\!\!$1$\!\!\!&\!\!\!$1$\!\!\!&\!\!\!$3$\!\!\!&\!\!\!$1$\!\!\!&\!\!\!$E\sst 6$\!\\[-.4mm]
\cline{2-12}
\!\!\!&\!\!\!$-1/3$\!\!\!&\!\!\!$-1$\!\!\!&\!\!\!$-1$\!\!\!&\!\!\!$-1$\!\!\!&\!\!\!$-1$\!\!\!&\!\!\!$1$\!\!\!&\!\!\!$1$\!\!\!&\!\!\!$1$\!\!\!&\!\!\!$3$\!\!\!&\!\!\!$1$\!\!\!&\!\!\!$E\sst 6$\!\\[-.4mm]
\cline{2-12}

\!\!\!&\!\!\!$-1/2$\!\!\!&\!\!\!$-1$\!\!\!&\!\!\!$-1$\!\!\!&\!\!\!$-1/2$\!\!\!&\!\!\!$0$\!\!\!&\!\!\!$1$\!\!\!&\!\!\!$0$\!\!\!&\!\!\!$i$\!\!\!&\!\!\!$2i$\!\!\!&\!\!\!$i-1$\!\!\!&\!\!\!$D\sst {2i+4}\yy B\sst 3^a\xx$\!\\[-.4mm]
\cline{2-12}
\!\!\!&\!\!\!$-1/2$\!\!\!&\!\!\!$-1$\!\!\!&\!\!\!$-1$\!\!\!&\!\!\!$0$\!\!\!&\!\!\!$1$\!\!\!&\!\!\!$2$\!\!\!&\!\!\!$0$\!\!\!&\!\!\!$i$\!\!\!&\!\!\!$2i$\!\!\!&\!\!\!$i-1$\!\!\!&\!\!\!$D\sst {2i+4}\yy B\sst 3\xx$\!\\[-.4mm]
\cline{2-12}
\!\!\!&\!\!\!$-1/2$\!\!\!&\!\!\!$-1$\!\!\!&\!\!\!$-1$\!\!\!&\!\!\!$-1$\!\!\!&\!\!\!$0$\!\!\!&\!\!\!$1$\!\!\!&\!\!\!$0$\!\!\!&\!\!\!$i$\!\!\!&\!\!\!$2i+1$\!\!\!&\!\!\!$i$\!\!\!&\!\!\!$D\sst {2i+4}$\!\\[-.4mm]
\cline{2-12}
\!\!\!&\!\!\!$-1/2$\!\!\!&\!\!\!$-1$\!\!\!&\!\!\!$-1$\!\!\!&\!\!\!$-1/2$\!\!\!&\!\!\!$0$\!\!\!&\!\!\!$1$\!\!\!&\!\!\!$0$\!\!\!&\!\!\!$i$\!\!\!&\!\!\!$2i+1$\!\!\!&\!\!\!$i$\!\!\!&\!\!\!$D\sst {2i+5}$\!\\[-.4mm]
\cline{2-12}
\!\!\!&\!\!\!$-1/2$\!\!\!&\!\!\!$-1$\!\!\!&\!\!\!$-1$\!\!\!&\!\!\!$0$\!\!\!&\!\!\!$1$\!\!\!&\!\!\!$2$\!\!\!&\!\!\!$0$\!\!\!&\!\!\!$i$\!\!\!&\!\!\!$2i+1$\!\!\!&\!\!\!$i$\!\!\!&\!\!\!$D\sst {2i+5}$\!\\[-.4mm]
\cline{2-12}
\!\!\!&\!\!\!$-1/2$\!\!\!&\!\!\!$-1$\!\!\!&\!\!\!$-1$\!\!\!&\!\!\!$-1$\!\!\!&\!\!\!$0$\!\!\!&\!\!\!$1$\!\!\!&\!\!\!$0$\!\!\!&\!\!\!$i+1$\!\!\!&\!\!\!$2i+2$\!\!\!&\!\!\!$i$\!\!\!&\!\!\!$D\sst {2i+5}$\!\\[-.4mm]
\cline{2-12}
\!\!\!&\!\!\!$-1/2$\!\!\!&\!\!\!$-1$\!\!\!&\!\!\!$-1$\!\!\!&\!\!\!$-1/2$\!\!\!&\!\!\!$0$\!\!\!&\!\!\!$1$\!\!\!&\!\!\!$0$\!\!\!&\!\!\!$(i-1)/2$\!\!\!&\!\!\!$i-1$\!\!\!&\!\!\!$i/2-1$\!\!\!&\!\!\!$B\sst {i+3}\yy G\sst 2\xx$\!\\[-.4mm]
\cline{2-12}
\!\!\!&\!\!\!$-1/2$\!\!\!&\!\!\!$-1$\!\!\!&\!\!\!$-1$\!\!\!&\!\!\!$0$\!\!\!&\!\!\!$1$\!\!\!&\!\!\!$2$\!\!\!&\!\!\!$0$\!\!\!&\!\!\!$(i-1)/2$\!\!\!&\!\!\!$i-1$\!\!\!&\!\!\!$i/2-1$\!\!\!&\!\!\!$B\sst {i+3}\yy G\sst 2\xx$\!\\[-.4mm]
\cline{2-12}
\!\!\!&\!\!\!$-1/2$\!\!\!&\!\!\!$-1$\!\!\!&\!\!\!$-1$\!\!\!&\!\!\!$-1$\!\!\!&\!\!\!$0$\!\!\!&\!\!\!$1$\!\!\!&\!\!\!$0$\!\!\!&\!\!\!$i/2$\!\!\!&\!\!\!$i$\!\!\!&\!\!\!$(i-1)/2$\!\!\!&\!\!\!$B\sst {i+3}\yy B\sst 3^a\xx$\!\\[-.4mm]
\cline{2-12}

\!\!\!&\!\!\!$-1$\!\!\!&\!\!\!$-1$\!\!\!&\!\!\!$-1$\!\!\!&\!\!\!$-1$\!\!\!&\!\!\!$-1$\!\!\!&\!\!\!$-1$\!\!\!&\!\!\!$-1$\!\!\!&\!\!\!$i-1$\!\!\!&\!\!\!$2i-1$\!\!\!&\!\!\!$i-1$\!\!\!&\!\!\!$C\sst i$\!\\[-.4mm]
\cline{2-12}

\!\!\!&\!\!\!$-1$\!\!\!&\!\!\!$-1$\!\!\!&\!\!\!$-1$\!\!\!&\!\!\!$z_3$\!\!\!&\!\!\!$z_4$\!\!\!&\!\!\!$z_5$\!\!\!&\!\!\!$z_6$\!\!\!&\!\!\!$z_7$\!\!\!&\!\!\!$z_8$\!\!\!&\!\!\!$z_9$\!\!\!&\!\!\!$A\sst {z_5+z_8+1}$
\!\\[-.4mm]

\hline\hline
\end{tabular}
\caption{Duals $\top^*$ of tops with $F_0^*$ one of the polygons $15,16$
of figure~\ref{fig:16pol}.}
\label{tab:results16}
\end{footnotesize}
\end{center}
\end{table}

\newpage
%\section*{}
%\newpage
%\suppressfloats
\appendix


\small

\begin{thebibliography}{11}

\def\I#1{{\it #1}}      \addtolength{\itemsep}{-4.5pt}  \small \vspace{-3mm}

\ifundefined{draftmode} \def\.#1 #2\>{\bibitem{#1}#2}
\else                   \def\.#1 #2\>{\bibitem{#1}\LLab{#1}#2}  \fi


\.Ba V.V. Batyrev, \emph{Dual Polyhedra and Mirror Symmetry for Calabi-Yau
Hypersurfaces in Toric Varieties},
J. Alg. Geom. \textbf{3} (1994) 493, alg-geom/9310003.
\>
\.CF P. Candelas, A. Font, \emph{Duality between the webs of heterotic and
type II vacua}, Nucl. Phys. \textbf{B511} (1998) 295, hep-th/9603170.
\>
\.egs E. Perevalov, H. Skarke, \emph{Enhanced Gauge Symmetry in Type II and
F-Theory Compactifications: Dynkin Diagrams From Polyhedra}, Nucl.Phys.
\textbf{B505} (1997) 679, hep-th/9704129.
\>
\.CPR1 P. Candelas, E. Perevalov, G. Rajesh, \emph{Comments on A,B,C Chains
of Heterotic and Type II Vacua}, Nucl. Phys. \textbf{B502} (1997) 594, 
hep-th/9703148.
\> 
\.CPR2 P. Candelas, E. Perevalov, G. Rajesh, \emph{F-theory duals of 
non-perturbative heterotic $E_8 \times E_8$ vacua in six dimensions}, 
Nucl. Phys. \textbf{B502} (1997) 613, hep-th/9606133.
\> 
\.CPR3 P. Candelas, E. Perevalov, G. Rajesh, \emph{Toric Geometry and 
Enhanced Gauge Symmetry of F-Theory/Heterotic Vacua}, Nucl. Phys. 
\textbf{B507} (1997) 445, hep-th/9704097.
\> 
\.CPR4 P. Candelas, E. Perevalov, G. Rajesh, \emph{Matter from Toric Geometry},
Nucl. Phys. \textbf{B519} (1998) 225, hep-th/9707049.
\> 
\.fst P. Candelas, H. Skarke, \emph{F-theory, $SO(32)$ and toric geometry},
Phys. Lett. \textbf{B413} (1997) 63, hep-th/9706226.
\> 
\.BM P. Berglund, P. Mayr, 
\emph{Heterotic String/F-theory Duality from Mirror Symmetry},
Adv. Theor. Math. Phys. \textbf{2} (1999) 1307, hep-th/9811217.
\> 
\.HLY Y. Hu, C.-H. Liu, S.-T. Yau, \emph{Toric morphisms and fibrations of 
toric Calabi-Yau hypersurfaces}, math.AG/0010082.
\>
\.fft M. Kreuzer, H. Skarke, \emph{Calabi-Yau Fourfolds and Toric Fibrations}, 
J. Geom. Phys. \textbf{26} (1998) 272, hep-th/9701175.
\> 
\.CHL S. Chaudhuri, G. Hockney, J. Lykken, \emph{Maximally Supersymmetric
String Theories in $D < 10$}, Phys.Rev.Lett. \textbf{75} (1995) 2264,
hep-th/9505054.
\> 
\.Fu W. Fulton, \emph{Introduction to Toric Varieties} 
(Princeton Univ. Press, Princeton, 1993). 
\> 
\.Oda   T. Oda, \I{Convex Bodies and Algebraic Geometry} (Springer, Berlin
        Heidelberg 1988).       \>
\.Cox  D. Cox, \I{The Homogeneous Coordinate Ring of a Toric Variety,}
        J. Alg. Geom. {\bf 4} (1995) 17, alg-geom/9210008. \>
\.sdt H. Skarke, \emph{String Dualities and Toric Geometry: An Introduction}, 
Chaos, Solitons \& Fractals \textbf{Vol. 10, No. 2-3} (1999) 543, 
hep-th/9806059. 
\> 
\.Kod K. Kodaira, \I{On Compact Analytic Surfaces II,} Ann. Math. {\bf 77}
        (1963) 563 .\>
\.BPV W. Barth, C. Peters, A. van de Ven, \emph{Compact Complex Surfaces}, 
Springer, 1984.
\>
\.GO P. Goddard, D. Olive, \emph{Kac-Moody and Virasoro Algebras in Relation 
to Quantum Physics}, Int. Journ. Mod. Phys. A, \textbf{Vol. 1, No. 2} (1986) 
303, (in \emph{Goddard, P. and Olive, D. (ed.): Kac-Moody Algebras and 
Virasoro Algebras: A Reprint Volume for Physicists} 8-119).
\> 
\.Ba85  V.V.Batyrev, \I{Higher-dimensional toric varieties with ample
        anticanonical class,} Moscow State Univ., Thesis, 1985 (Russian)\>
\.Koe  R.J.Koelman, \I{The number of moduli of families of curves on toric
        varieties,} Katholieke Universiteit Nijmegen, Thesis, 1990
        \>
\.crp M. Kreuzer, H. Skarke, \emph{On the Classification of Reflexive 
Polyhedra}, Commun. Math. Phys. \textbf{185} (1997) 495, hep-th/9512204.
\> 
\.wtc H. Skarke, \emph{Weight Systems for Toric Calabi-Yau Varieties and 
Reflexivity of Newton Polyhedra}, Mod. Phys. Lett. \textbf{A11} (1996) 1637, 
alg-geom/9603007.
\> 
\.rwf M. Kreuzer, H. Skarke, \emph{Reflexive Polyhedra, weights and toric 
Calabi-Yau fibration}, Rev. Math. Phys. \textbf{14} (2002) 343, 
math.ag/0001106.
\> 
\.c3d M. Kreuzer, H. Skarke, \emph{Classification of Reflexive Polyhedra in 
Three Dimensions}, Adv. Theor. Math. Phys. \textbf{2} (1998) 853, 
hep-th/9805190.
\> 
\.c4d M. Kreuzer, H. Skarke, \emph{Complete classification of reflexive
polyhedra in four dimensions}, Adv. Theor. Math. Phys. \textbf{4} (2002) 1209, 
hep-th/0002240.
\> 
\.BIKMSV M. Bershadsky, K. Intriligator, S. Kachru, D. R. Morrison, V. Sadov,
        C. Vafa, \I{Geometric Singularities and Enhanced Gauge Symmetries,}
        \npb 481 (1996) 215, hep-th/9605200. \>
\.AG   P.S. Aspinwall and M. Gross, \I{The SO(32) Heterotic String on a $K3$
        Surface,} Phys. Lett. {\bf B387} (1996) 735, hep-th/9605131. \>
\.CP S. Chaudhuri, J. Polchinski, \emph{Moduli Space of CHL Strings}, 
Phys.Rev. \textbf{D52} (1995) 7168, hep-th/9506048.
\> 
\.SS J. Schwarz, A. Sen, \emph{Type IIA Dual of the Six-Dimensional CHL 
Compactification}, Phys.Lett. \textbf{B357} (1995) 323, hep-th/9507027.
\> 
\.BPS M. Bershadsky, T. Pantev, V. Sadov, 
\emph{F-Theory with Quantized Fluxes}, 
Adv. Theor. Math. Phys. \textbf{3} (1999) 727, hep-th/9805056.
\> 
\.Pa J. Park, \emph{Orientifold and F-Theory Duals of CHL Strings}, Phys. 
Lett. \textbf{B418} (1998) 91, hep-th/9611119.
\> 
\.BKMT P. Berglund, A. Klemm, P. Mayr, S. Theisen, 
\emph{On Type IIB Vacua With Varying Coupling Constant}, 
Nucl. Phys. \textbf{B558} (1999) 178, hep-th/9805189.
\> 
\.LMST W. Lerche, R. Minasian, C. Schweigert, S. Theisen, \emph{A Note on the 
Geometry of CHL Strings}, Phys.Lett. \textbf{B424} (1998) 53, hep-th/9711104.
\> 
\.CL S. Chaudhuri, D.A. Lowe, \emph{Type IIA-Heterotic Duals With Maximal 
Supersymmetry}, \npb 459 (1996) 113, hep-th/9508144.
\>
\.KKO S. Kachru, A. Klemm, Y.Oz, \emph{Calabi-Yau Duals for CHL Strings}, 
Nucl.Phys. \textbf{B521} (1998) 58, hep-th/9712035.
\> 
\del
\.FMW R. Friedman, J. Morgan, E. Witten, \emph{Vector Bundles and F-theory}, 
Commun. Math. Phys. \textbf{187} (1997) 679, hep-th/9701162.
\> 
\.Sc C. Schweigert, \emph{On moduli spaces of flat connections with non-simply 
connected structure group},  Nucl.Phys. \textbf{B492} (1997) 743, 
hep-th/9611092.
\> 
\.Wi E. Witten, \emph{Toroidal Compactifications Without Vector Structures}, 
JHEP 9802:006 (1998), hep-th/9712028.
\>
\enddel



\end{thebibliography}
\bye